# Estimation of Linear models
# from Coarsened Observations[1]

## A Method of Moments Approach

**Bernard M.S. van Praag\*, J. Peter Hop†\*\*, William H. Greene\*\*\***

\*Tinbergen Institute, University of Amsterdam, The Netherlands
\*\*Independent, The Netherlands(†august,2,2024)
\*\*\*University of South Florida, U.S.
Corresponding author: b.m.s.vanpraag@uva.nl
Monday, 29 october 2024

---

[1] We are grateful for helpful remarks by the referees.

**Abstract**

**Estimation of Linear models
from Coarsened Observations**
A Method of Moments Approach

In the last few decades, the study of ordinal data in which the variable of interest is not exactly observed but only known to be in a specific ordinal category has become important. In Psychometrics such variables are analysed under the heading of item response models (IRM). In Econometrics, subjective well-being (SWB) and self-assessed health (SAH) studies, and in marketing research, Ordered Probit, Ordered Logit, and Interval Regression models are common research platforms. To emphasize that the problem is not specific to a specific discipline we will use the neutral term coarsened observation. For single-equation models estimation of the latent linear model by Maximum Likelihood (ML) is routine. But, for higher -dimensional multivariate models it is computationally cumbersome as estimation requires the evaluation of multivariate normal distribution functions on a large scale. Our proposed alternative estimation method, based on the Generalized Method of Moments (GMM), circumvents this multivariate integration problem. The method is based on the assumed zero correlations between explanatory variables and generalized residuals. This is more general than ML but coincides with ML if the error distribution is multivariate normal. It can be implemented by repeated application of standard techniques. GMM provides a simpler and faster approach than the usual ML approach. It is applicable to multiple -equation models with $K$ -dimensional error correlation matrices and $J_k$ response categories for the $k^{th}$ equation. It also yields a simple method to estimate polyserial and polychoric correlations. Comparison of our method with the outcomes of the Stata ML procedure `cmp` yields estimates that are not statistically different, while estimation by our method requires only a fraction of the computing time.

## 1. Introduction.[2]

The statistical tools of empirical Psychometrics, Econometrics, Political Science, and many other empirical sciences including marketing analysis, agriculture, health, and medical statistics, find their origin in the linear regression model. The idea is that a random phenomenon $Y$ can be *predicted* by variables $X$ in the sense that $Y \approx f(X; \beta)$, where $\beta$ is a parameter vector. Later, the same model is applied to *explain* the phenomenon $Y$, as caused by the variables $X$. In the econometrics literature since the 1960s this resulted in a host of different models, described in textbooks such as Greene (2018) and Cameron and Trivedi(2005).) In psychometrics there was a similar development known as the Structural Equation Model (SEM) (Duncan (1975), Hayduk, (1987), Bollen (1989), Jöreskog(1973).) (For software, see, for example, Narayanan (2012).) The main idea behind the modeling approach is that the phenomenon $Y$ to be studied has a conditional mean function that depends on other variables $X_1, ..., X_M$, say, $E(Y | X = x) = f(x; \beta)$, where $f(.)$ is a continuous and differentiable function and $\beta$ is a set of parameters. In practice the model is usually taken to be linear, that is, $E[Y | X = x] = \beta_0 + \beta_1 x_1 + ... + \beta_M x_M$, which produces the linear regression model. In the traditional approach, the variable $Y$ is continuous and directly observable. In economics the model approach (a first influential introduction is by Hood and Koopmans(1953)) is used to describe dependencies between economic variables like, e.g., purchase intentions as a function of income, prices, education, family size, and age. In marketing the modeling approach is used to develop and assess the effects of advertising, prices, promotions, etc. (see e.g. Fok (2017)). In medical statistics a model is used to evaluate the response to diagnostic tests. These are a few examples to illustrate the pervasiveness of the regression model approach in empirical sciences.

The one-dimensional observation $Y_i$ may be replaced by a $K$-dimensional vector, $Y_i = (Y_{i1}, ..., Y_{iK})$ and the function $f(.)$ by a $K$-vector function, $f = (f_1, ..., f_K)$.

---

[2] Capitals will be used for random variables, vectors, and matrices. We denote the zero-vector by o=(0,...,0). Disturbance terms will be in Greek letters. Roman letters will be used for constants and realizations of random variables. Matrices will be denoted by capitals as well to conform to the traditional regression formulas. Indexes will be suppressed where the interpretation will be clear from the context. The ML-estimator of a parameter vector $\theta$ is denoted by $\hat{\theta}$. The corresponding estimator from coarsened data is denoted by $\hat{\tilde{\theta}}$. An overbar above a symbol denotes a sample average.

In practice, the parameters of interest in such a model are estimated from a set of $N$ observations $\left( \{Y_i, X_i\} \right)_{i=1}^{N}$. Randomness enters the observed outcomes through the difference between the individual response, $Y_i$, and the conditional expectation for individual observation $i$, $E(Y_i | X_i = x_i) = f(x_i)$. Hence, we introduce an 'error,' $\varepsilon_i$, or a $K$-vector of errors, $\varepsilon_i = (\varepsilon_{i1}, ..., \varepsilon_{iK})$, which is defined for each response as $\varepsilon_i \stackrel{def}{=} Y_i - f(X_i; \beta)$. The error represents the aggregate of other possible, unobserved variables as well as the randomness of individual behavior. The random term is generated by a mean zero process that operates stochastically independently of $X$. If the model is linear, we denote the discrepancy as the residual $\varepsilon \stackrel{def}{=} Y - X'\beta$. This approach is applicable when the dependent variable(s) $Y$ and the explanatory variable(s) $X$ is (are) cardinal, i.e., are expressed in observable numerical values.

However, in many practical cases the dependent variable $Y$ is coarsened; it is only observable in terms of ordinal categories on a preference scale, such as subjective health status, well-being, , reported as 'bad' or 'good', or 'like' or 'dislike' ,or "poor, fair, good, very good, excellent," or some ordinal ranking from one to five where a cardinal interpretation becomes dubious. Although the observations take place in a coarsened mode, we must interpret the discrete answers as reflecting a latent variable $Y$, the range of which is a continuum. In that situation, we say the observations are 'coarsened' or condensed into a set of $J$ adjacent intervals $\left\{ (v_{j-1}, v_j] \right\}_{j=1}^{J} = \left\{ S_j \right\}_{j=1}^{J}$ on the real axis. For an individual $i$, the latent observation is $Y_i \in S_{j_i}$, if the realized observation is $j_i$. Hence, we see that the estimation of the model is complicated by two factors. First, there is the statistical problem that there is always a random error term involved. Second, there is the additional observation problem that the continuum of observations is coarsened or condensed (Maris (1995)) and mapped on a discrete event space $\left\{ j = 1, ..., J \right\}$. We will call such data Ordinally Coarsened (OC).

Since about 1934 in bioassay studies (Bliss (1934), Finney (1947, 1971)), and in Economics, Sociology and Political Science, much later in the 1960s and 1970s researchers realized that many variables of interest have an ordinal coarsened character. For instance, a question on self-assessed health status may be responded to with ordered labels varying from 'very healthy,' to 'very unhealthy'. In modern datasets, especially survey data, such verbal evaluations are abundant. The *World Happiness Report* (*Oxford University, 2024*) is a notable example. Such a coarsening is nearly always dictated by the fact that respondents are unable to quantify their answers directly on a numerical continuous scale but only in terms of a few ordered verbally described categories. These values are not expressed in numbers but in ordinal qualitative terms. In Psychometrics popular item response theory offers many examples (see Wiberg et al. (2019)). To stress that the statistical problems are not specific for one discipline we will also speak of coarsened data.

In psychometric Structural Equations models (SEM) things may be further complicated by the fact that the explanatory $X$ variables are sometimes coarsened as well (cf. Jöreskog (1973)). In this paper we will assume that the explanatory $X$ variables are either directly observable on a continuous scale, or coarsened in ordered classes labelled 1,2,.., or by dichotomous dummy variables, taking the value zero or one.

A specific case of coarsened events is that of psychological testing (item -response theory IRT) and that of multiple-choice tests used in exams. Say, the exam or test consists of $K$ items with each $J$ ordered response categories. Then the test may be described by *K* item scores $Y_1, ..., Y_K$. The response to each separate item may be dichotomous (e.g. correct /false) or polychotomous (correct, not wholly correct,..., false).

The probability of the response on item $k$ is $p_{ikj} = F(b_k; \theta_i)$, where $b_k$ stands for difficulty of the item $k$ and $\theta_i$ for the ability of respondent *i.* From the joint probability of the $K$ responses by individual *i* we try to estimate the latent ability $\theta_i$ (e.g., IQ) of the individual by maximizing per individual *i* the joint probability $\prod_{k=1}^{K} F(b_k; \theta_i)$, with respect to $\theta_i$. The ability $\theta_i$ may also depend on (or be explained by) individual observable characteristics $X_i = (X_{i1}, ..., X_{iM})$, say $\theta_i = X_i'\beta + D_i\beta_{0i}$, where the dummy variable $D_i$ equals one for individual *i* and zero for others and $\beta_{0i}$ is the ability of individual *i*. The term $X_i'\beta$ gives then the part of ability or intelligence that can be explained by e.g. education, genetic factors, health, income, etc. , while the individual parameter $\beta_{0i}$ may be identified as the unexplainable truly individual ability component.

We notice that estimation of $\beta_{0i}$ requires that *K>1* and preferably considerably larger than one,

The method developed below may be used to estimate these $\beta$'s and $\theta_i$'s. The response probability $F(b_k;\theta_i)$ of a separate item score is frequently assumed to be described by a normal or logistic distribution function. Estimation of $\theta_i$ is then rather easy. But intuition tells us that mostly the item scores on different items by a respondent will be correlated. This is almost unavoidable if the response behavior depends on one $\theta_i$. In that case the ML- estimation is mostly difficult since estimation will require the evaluation of many multivariate integrals. The correlation makes that the multivariate integrals can not be reduced to products of one-dimensional integrals. The method proposed by us will avoid this problem.

For directly observed cardinal data, ordinary least squares (OLS) is usually the default estimator of choice[3]. There are many extensions of the OLS estimator that are used in nonstandard cases, such as nonzero covariances across observations. A familiar alternative to OLS is Generalized Least Squares (GLS), in which the disturbances of the $K$ observations per observation unit $i$ are heteroskedastic or correlated. Then an unknown error-covariance matrix has to be estimated as well. If this is feasible, we call the method Feasible Generalized Least Squares (FGLS). Another well-known example is Seemingly Unrelated Regressions (SUR), where $K$ response variables are explained by $K$ equations where the errors are correlated. We refer to well-known econometric textbooks such as Amemiya(1985),(Cameron and Trivedi (2005), Greene (2018), and Verbeek (2017) for elaborate descriptions. We note that in modern work these estimations are usually based on the assumption of a known error distribution, usually multivariate normal, leading to Maximum Likelihood (ML) estimation. In the early received literature, least squares was not explicitly based on an underlying error distribution, but rather on minimization of a Sum of Squared Residuals (SSR) that led to unbiased estimation of the regression coefficients. Later it was found that SSR-minimization and ML-estimation led to the same estimator when the errors are normally distributed. (The normality assumption was also used to motivate certain inference procedures.) The common counterpart for the linear regression model in case of coarsening of $Y$ is the Ordered Probit or Ordered Logit model (OP or OL). In the literature it is frequently called Probit or Logit Regression. (See e.g. McElvey and Zavoina (1975)).

---

[3] The original least squares method is due to Gauss (1809) and Legendre (1804) A.M. Legendre. *Nouvelles méthodes pour la détermination des orbites des comètes*, Firmin Didot, Paris, 1805. "Sur la Méthode des moindres quarrés". See, also, Stigler (1981) for a historical survey.

In this paper we will develop a novel approach to coarsened data, called Feasible Multivariate Ordered Probit (FMOP), where the errors are suspected of being correlated, as it is the case in, e.g., item response models, in panel data, regional data or customer satisfaction data. It follows the analogy principle as formulated by Goldberger (1968, 1991) and Manski (1988), based on the method of moments (MoM).

The established way to treat such models with coarsened correlated observations in psychometrics or econometrics, and other empirical applications is by maximum likelihood (ML) estimation (see McFadden (1989), Hajivassiliou and McFadden (1998)), where the likelihoods per observation include $K$-variate normal integrals (instead of densities). Those integrals are generally estimated by simulation such as by the Geweke-Hajivasilliou-Kean (GHK)-algorithm. Theoretically, this is a straightforward application of ML theory. However, the practical problem is that the evaluation of those integrals by simulation may be quite cumbersome and time-consuming.

Geweke (1989), McFadden (1989), Keane (1994), Hajivassiliou and McFadden (1998), Cappellari and Jenkins (2003), Mullahy (2016), and others developed a multivariate probit estimator. Progress has also been made based on the simulated moments approach (McFadden (1989)), using the Gibbs sampler method (see Geman and Geman (1984), and Casella and George (1992)). An interesting historical survey on the (logit and) probit method is found in Cramer (2004). See also Hensher, Rose, and Greene (2015). Roodman (2007, 2020) developed a flexible working Stata estimation procedure (`cmp`) based on the GHK-simulator.

Independently, scholars in Psychometrics expanded a vast literature on IRT-models yielding different tools of analysis, inspired by the differences between the subject matters between disciplines (See Bollen (1989)). It is surprising that both psychometricians and econometricians are working on essentially the same methodological problems, but mostly without taking much notice of each other's literature. A rare exception is the econometrician Goldberger (1971) who explicitly recognized the commonalities between Econometrics and Psychometrics.

Our approach, based on sample moments, does not need the evaluation of likelihoods,

i.e. ,multi-dimensional probability integrals or simulated moments. In this paper we assume multivariate normal error distributions for $\varepsilon$ in line with the established practice. Data on $X$ are assumed to be generated by a random process that is statistically (and functionally) independent of that which generates $\varepsilon$. The important element is that the observed values of $X_i$ convey no information about the errors $\varepsilon_i$, a property identified by econometricians as 'strict exogeneity.' It is the property that $E[\varepsilon|X] = 0$ for every $X$. It follows that $\text{Cov}(X,\varepsilon) = \text{Cov}(X, E[\varepsilon|X]) = 0$ as well. Econometricians call this condition strict exogeneity. It implies the same zero- covariance property. Data on $X$ are assumed to be "well behaved", meaning that in any random sample, the sample covariance matrix $(1/N)\sum_{i=1}^{N} X_i' X_i \overset{def}{=} \hat{\Sigma}_X$ is always finite and positive definite. (Regularity conditions on $X$ such as that the influence of any individual $X_i$ in $(1/N)\sum_{i=1}^{N} X_i' X_i$ vanishes as $N$ increases are also assumed.) Nothing further is assumed at this point about the distribution of $X_i$, e.g., normality, discreteness, etc.

The theoretical model, that is, the data-generating mechanism, mimics the classical linear statistical models. The substantive difference between exact and coarsened data is in the mode of observation of the dependent variable $Y$. In the classical framework the dependent variable $Y$ is directly observed, while in the Ordered Coarsened (OC) - observation mode, the dependent variable $Y$ is only observed to be in one of $J$ intervals $S_j = (\nu_{j-1}, \nu_j]$, where the cut points $\nu_{j-1}$ and $\nu_j$ are unknown parameters to be estimated. If $Y$ is a $K$-vector, $\nu_{j-1}$ and $\nu_j$ are $K$-vectors and $S_j = (\nu_{j-1}, \nu_j]$ is a block in $R^K$. Since the cut points $\nu_{j-1}$ and $\nu_j$ are unknown, it follows that the unit of measurement of $Y$ is unidentified. The usual identification is secured by setting the error variances equal to one; $\sigma_k^2 = 1 \ (k = 1, ..., K)$ .. [4]

The structure of this paper is as follows. In Section 2 we outline the basic probabilistic model in the presence of coarsening of the dependent variables. In Section 3 we develop the estimation method for a $K$-equations model based on the Ordered Coarsened data model. In IRT-models this is equivalent to $K$ item responses per respondent. We call the method Feasible Multivariate Ordered Probit (FMOP). We call it Seemingly Unrelated Ordered Probit (SUOP), when we have a $K$-equation model with one observation $Y_i = (Y_{i1}, ..., Y_{iK})$ per observation unit $i$. Of course, mixtures of FMOP and SUOP are possible as well. Instead of differentiating high-dimensional log-likelihoods, with the likelihoods being multi-dimensional integrals, with respect to the unknown parameters $\theta = (\beta, \nu, \Sigma)$, we derive sample moment conditions $\hat{\bar{g}}(\theta) = 0$ from the coarsened data that are analogues of the likelihood equations, $\hat{g}(\theta) = 0$ for direct observations. We then estimate $\theta$ from the equation set $\hat{\bar{g}}(\theta) = 0$. We find that $\hat{\bar{g}}(\theta)$ and $\hat{g}(\theta)$ converge to the same probability limit for all values of $\theta$. The estimation of $\theta$ from $\hat{\bar{g}}(\theta) = 0$ takes place by repeatedly applying the generalized method of moments (GMM), (Hansen (1982)). We note in passing, our approach employs elements of the EM algorithm (Dempster, Laird and Rubin (1977). In Section 4 we demonstrate how to estimate polychoric correlations from SUOP-estimates. Further, we generalize the concept of the coefficient of determination, $R^2$, to $K > 1$. In Section 5a we apply FMOP to an employment equation over a five-year panel-dataset from the German SOEP dataset. In Section 5b we apply SUOP to a block of eight satisfaction questions extracted from the German SOEP data. In order to get more insight into the stability of the method, in Section 5c. we do some experiments on a simulated data set. We use a recent update of the Stata procedure `cmp` as our benchmark to compare our alternative approach with the ML approach.

---

[4] The important issue of whether the coarsened sample data $(\overline{Y}_i, X_i)$ contain sufficient information to identify estimators of $\beta$ and the unknown cut-points in $S_j$ is considered in Greene and Hensher (2010) among others.

We find that the estimation results of regression coefficients and their standard errors do not differ substantially between the two methods. Where the established method may need hours, our method takes only minutes. In Section 6 we provide some concluding remarks. In the Appendix we propose an easy intuitively appealing method to estimate the latent full error-covariance matrix, after the regression coefficients, $\beta$, have been estimated using the coarsened data.

## 2. Regression in the population space

Regression analysis often begins with assumptions about the distribution of the observed data. Each estimation procedure on a sample may be seen as a reflection of a similar procedure in the population. For convenience but without loss of generality we assume $E(X) = 0$, $E(Y) = 0$. The model of interest for equation $k$ of $K$ equations is

$$Y_k \mid x = \sum_{m=1}^{M} x_{m,k} \beta_{m,k} + \varepsilon_k = x'_k \beta_k + \varepsilon_k, \ \ k = 1, ..., K \tag{2.1}$$

Where the model design might call for differences across equations, they can be accommodated by suitable zero restrictions on the coefficients in $\beta_k$. For convenience, we ignore constant terms by setting $E(X) = 0$. The expectation of outcome or response $Y_k$ is conditioned or co-determined by explanatory variables /stimuli $X_m$, $(m = 1, ..., M)$ assuming values $x_m$. There are $M$ variables $X_m$, that are generated by a strictly exogenous process. The error vector $\varepsilon$ derives from observation - specific variation around the theoretical conditional mean. We have $\varepsilon_k = Y_k - X'\beta_k$. Denoting the observations by $y_{i,k}$ the observed residual is defined as $e_{i,k} = y_{i,k} - x'_i \beta_k$. Deviations of observations $i = 1, ..., N$ of $Y_{i,k}$ from the conditional mean result from the presence of unobserved elements that enter the data generating process, for example, variation across individuals in the self-assessment of health or well-being. Random elements are assumed to be generated by a zero mean, finite variance process; if $\varepsilon_k = Y_k - X'\beta_k$, it follows that $E[\varepsilon] = 0$ and $\text{Var}[\varepsilon]$ is finite. We assume that the process that generates $X$ is stochastically independent of that of $\varepsilon$. This implies $E[X\varepsilon] = 0$. Substitution yields the familiar regression equations

$$E\left[ X'_i (Y_{i,k} - X'_i \beta_k) \right] = 0, \ k = 1, ..., K \ , \tag{2.2}$$

where $X'_i$ is a $(M \times K)$-matrix and $X'_i (Y_{i,k} - X'_i \beta_k)$ a $K$-vector.

If this holds for all $i$, then we have

$$\sum_{i=1}^{N} E\left[ X'_i (Y_{i,k} - X'_i \beta_k) \right] = 0 \ \forall k, m \tag{2.3}$$

from which the regression coefficients $\beta$ can be solved.

We define the functions $g_{m,k}(\beta) = \sum_{i=1}^{N} E\left[(X'_{i,k,m})(Y_{i,k} - X'_i\beta_k)\right]$. The equation system (2.3) in $\beta$ is then shortly written as $g(\beta) = 0$.

Result (2.3) identifies the slopes $\beta$ of the conditional mean function. The zero conditional mean result in (2.2) motivates least squares without reference to minimizing the mean squared error prediction of $Y|X$ or minimum variance linear unbiased estimation of $\beta$.

If the error covariance matrix $\Sigma = E(\varepsilon\varepsilon')$ is not the identity matrix, we may want to correct for the unequal variances and correlations, and we weigh the observations by $\Sigma^{-1} = \left[E\left(Y - X'\beta\right)\left(Y - X'\beta\right)'\right]^{-1}$, producing

$$g_{m,k}(\beta) = \sum_{i=1}^{N} E\left[(X'_{i,m})\Sigma^{-1}(Y_{i,k} - X'_i\beta_k)\right].$$

By weighting the $K$ observations per individual $i$ by $\Sigma^{-1}$ variances are standardized, and the error correlations accounted for. In that case, according to the Aitken theorem the covariance matrix of the estimator $\hat{\beta}$ is minimized.

We have motivated least squares through the moment equations (2.2). We see that we can interpret these conditions (2.2) as first-order conditions for minimizing the expectation of the squared residuals $S = E(\varepsilon'\Sigma^{-1}\varepsilon) = E\left(Y - X'\beta\right)'\Sigma^{-1}\left(Y - X'\beta\right)$.

We do not need to specify the probability distribution of $X$ and $Y$. We do assume well-behaved data generating processes, which will include a finite, positive variance of $\varepsilon$ and a finite positive covariance matrix, $\text{Var}[X,Y]$. If the marginal probability distribution of $\varepsilon$ is multivariate normal, the regression estimator is Maximum Likelihood. We call $X'\beta$ the structural part of the model and $\varepsilon$ the disturbance, where $\beta$ is the $(M \times K)$ – matrix with columns $\beta_1', ..., \beta_K'$.

If the columns of the matrix $\beta$ are identical, this is the typical setting for longitudinal and panel data. If the coefficients vary by response setting $k$, $\beta_k$ we have the situation of $K$ different model equations.

*Regression for Coarsened observations*

We call an observation $Y \in R$ coarsened if it is not observed directly, but only as belonging to one of the $J$ intervals, $\{(\nu_{j-1}, \nu_j]\}_{j=1}^{J} = \{S_j\}_{j=1}^{J}$ (The leftmost and rightmost terminals are infinite). These intervals constitute the class $\mathbb{C}$ of observable events. We will call $\mathbb{C}$ the observation grid. This is generalized to $K$-dimensional observations by replacing the $J$ observed intervals $(\nu_{j-1}, \nu_j]$ by $K$-dimensional blocks

$$(\nu_{j-1}, \nu_j] = \left[ (\nu_{1,j_1-1}, \nu_{1,j_1}], ..., (\nu_{K,j_K-1}, \nu_{K,j_K}] \right] = S_j$$

where the one-dimensional random observations $\dot{j}_i$ are replaced by the index vectors $j_i = (j_{i1}, ..., j_{iK})$. $\mathbb{C}$ stands for a partition in $R^K$ consisting of $J^K$ adjacent blocks. (We take $J_1 = ... = J_K \overset{def}{=} J$ for convenience, but equality is not necessary). We denote the $\mathbb{C}$-coarsened event space by $\Omega_{\mathbb{C}}$. We may then define the corresponding coarsened probability measure $P_{\mathbb{C}}$ on $\mathbb{C}$ by $P_{\mathbb{C}}$. We will call $\mathbb{C}$ the $K$-dimensional observation grid.

Coarsening of $Y$ implies that we do not directly observe $Y = y$, but the event $Y \in S_j$, and more explicitly for a $K$-vector $Y$ that $\nu_{1,j_1-1} < Y_1 \leq \nu_{1,j_1}, ..., \nu_{K,j_K-1} < Y_K \leq \nu_{K,j_K}$. It follows that for given $X = x$ and $j = (j_1, ..., j_K)$ there holds for the error vector

$$\nu_{1,j_1-1} - x_1'\beta_1 < \varepsilon_1 \leq \nu_{1,j_1} - x'\beta_1, ..., \nu_{K,j_K-1} - x_K'\beta_K < \varepsilon_K \leq \nu_{K,j_K} - x_K'\beta_K,$$

which we denote shortly as $\varepsilon \in (S_j - x\beta)$. We denote the marginal probability as $p_{k,j} = P(\nu_{1,j-1} - x\beta_k < \varepsilon_k \leq \nu_{1,j} - x\beta_k)$. We define the generalized residual as $\overline{\varepsilon} \overset{def}{=} E(\varepsilon | \varepsilon \in (S_j - x\beta))$. It is a random $K$-vector defined on the blocks $(\nu_{j-1} - x_i\beta, \nu_j - x_i\beta]$. These blocks constitute individual $i's$ individualized observation grid $\mathbb{C}$. The grid $\mathbb{C}_i$ for observation unit $i$ depends on $x_i'\beta$. However, for any given value $x$ of $X$ and any grid $\mathbb{C}$ $(x\beta)$ we have $\sum_{j=1}^{J} p_{k,j} \overline{\varepsilon}_{k,j} \overset{def}{=} E_{\mathbb{C}}(\overline{\varepsilon}_k) = E(\varepsilon) = 0$ according to the Law of Iterated

Expectations(LIE). Then it follows that

$$x_{i,k}.E_{\mathbb{C}}(\overline{\varepsilon}_{i,k}) = 0, \ \forall i, k \tag{2.3}$$

and

$$(1/N)\sum_{i=1}^{N} E_{X_{i,k}}(X_{i,k}.\overline{\varepsilon}_{i,k}) = 0, \ k = 1,...,K \tag{2.4}$$

This is the coarsened analogue of (2,2). We define the function $\overline{g}_k(\beta) = (1/N)\sum_{i=1}^{N} E\left(X_i(\overline{\varepsilon}_{i,k})\right)$

. The equation system (2.4) is then shortly written as $\overline{g}(\beta) = o$. If the error covariance matrix is not the identity matrix, we may want to correct for the unequal variances and correlations, and we write $\overline{g}(\beta) = \sum_{i=1}^{N} X_i \overline{\Sigma}^{-1} E(\overline{\varepsilon}_{i,k})$, where $\overline{\Sigma} = \text{Var}(\overline{\varepsilon})$.

Finally, we have for the two functions

$$g(\beta) \equiv \overline{g}(\beta) \tag{2.5}$$

This holds for all values of $\beta$, not only for the zero roots of (2.2) and (2.4). (2.5) holds since for any $x$ - value $xE(\varepsilon) = xE(\overline{\varepsilon})$.

## 3. Large Sample Results for Regression

The Law of Large Numbers states that under standard regularity conditions, sample

moments converge in probability to their population counterparts as the number $N$ of

observations grows large. Slutsky's theorem says that continuous and differentiable

functions of random sample moments converge in probability to those functions of the

population counterparts as $N$ grows large, implying that the population functions

$g(\beta, \Sigma)$ are consistently estimated by filling in the corresponding sample moments.

When we want to get its (large-)sample estimator $\hat{g}(\beta, \Sigma)$ we replace the

expectations in (2.2) by the corresponding sample moment conditions and we get

$$\frac{1}{N}\sum_{i=1}^{N} X_i' \hat{\Sigma}^{-1} Y_i - \left(\frac{1}{N}\sum_{i=1}^{N} X_i' \hat{\Sigma}^{-1} X_i'\right)\hat{\beta} \overset{def}{=} \hat{g}(\hat{\beta}, \hat{\Sigma}) = 0, \tag{3.1}$$

where $\hat{\Sigma} = \frac{1}{N}\sum_{i=1}^{N}(Y_i - \beta'X_i)(Y_i - \beta'X_i)'$. Solution of (3.1) with respect to the regression coefficients $\beta$ yields

$$\hat{\beta} = \left[\frac{1}{N}\sum_{i=1}^{N}(X_i'\hat{\Sigma}^{-1}X_i)\right]^{-1}\frac{1}{N}\sum_{i=1}^{N}(X_i'\hat{\Sigma}^{-1}Y_i).$$ (3.2).

This is the well- known OLS- estimator.

Estimation of the asymptotic covariance matrix Asy.Var$[\hat{\beta}]$ is usually understood to mean under the condition that $X$ equals the sample data $x$. Then the well-known template is

$$\text{Est.Asy.Var}[\hat{\beta}] = \frac{1}{N}\left[\frac{1}{N}\sum_{i=1}^{N}(X_i'\hat{\Sigma}^{-1}X_i)\right]^{-1}$$ (3.3)

*Estimation from ordered coarsened data.*

Let us now consider the coarsened analogue. The events are elements of the observation grid $\mathbb{C}$. The corresponding coarsened probability measure is $P_\mathbb{C}$. If we would follow the conditional ML-strategy, the information to be maximized is. $E_\mathbb{C}$ (ln $P_\mathbb{C}$)=

$$\sum_{i=1}^{N}\sum_{j=1}^{J} \ P_\mathbb{C} \ (Y_i \in S_j \mid x_i)\ln \ P_\mathbb{C} \ (Y_i \in S_j \mid x_i)\ .$$

The problem is here the evaluation of the $P_\mathbb{C}$ $(Y_i \in S_j \mid x_i)$, being multivariate integrals over rectangular blocks in $R^K$. We can evaluate $P_\mathbb{C}$ $(Y_i \in S_j \mid x_i)$ by its sample analogue, but this entails the evaluation of a multitude of $K$-dimensional integrals, making this procedure very cumbersome, albeit not impossible (see Roodman (2011)).

A much easier way is by making a detour and evaluating the coarsened analogue of the condition (3.1). Let $j_i$ be the $K$-dimensional response by individual *i*. We notice that (the $K$-dimensional) $Y_i \in S_{j_i} \mid x_i$ implies $\varepsilon_i \in (v_{i,j_i-1} - x_i'B, v_{i,j_i} - x_i'B]$. In this way to each observation unit $i$ is assigned its own observation grid $\mathbb{C}_i$ depending on $x_i'B$.

We define the $K$-vector of the generalized residuals

$$\overline{\varepsilon}_i = E\left[\varepsilon_i \Big| \varepsilon_i \in (v_{i,j_i-1} - x_i'B, \ v_{i,j_i} - x_i'B], X_i = x_i\right].$$

The grid $\mathbb{C}_i$ over which the expectation is taken at the LHS in (3.1) is now a grid in the space $R^{M+K}$ where the first $X$-coordinates are directly observed while the $K$ $\varepsilon_i$-coordinates are coarsened by $\mathbb{C}_i$. Summing over the observations we get

$$\frac{1}{N}\sum_{i=1}^{N} E\left[x_i'\Sigma^{-1}\varepsilon_i \mid x\right] = \frac{1}{N}\sum_{i=1}^{N} E_{\mathbb{C}^i}\left[x_i'\overline{\Sigma}^{-1}\overline{\varepsilon}_i \mid x\right] \qquad (3.4)$$

Then (3.4) may be summarized as the identity

$$g(\beta,\Sigma \mid x) \equiv \overline{g}(\beta,\overline{\Sigma} \mid x) \qquad (3.5)$$

This implies that the equations $\overline{g}(\beta,\overline{\Sigma} \mid x) = 0$ and $g(\beta,\Sigma \mid x) = 0$ have the same roots $\beta$. The vector function $\overline{g}(\beta,\Sigma)$ may be interpreted as the vector of derivatives of a criterion function like a log-likelihood or the sum of squared residuals with respect to $\beta$. The simplest criterion function with $\overline{g}(\beta,\Sigma)$ as gradient vector[5] is

$$\overline{S} = \frac{1}{N}\sum_{i=1}^{N} E(\overline{\varepsilon}'\,\overline{\Sigma}^{-1}\overline{\varepsilon} \mid X = x_i) =$$

$$= \frac{1}{N}\sum_{i=1}^{N} E\left[\varepsilon_i \Big| \varepsilon_i \in (v_{i,j-1} - x_i'\beta,\ v_{i,j} - x_i'\beta], X = x_i\right]' \overline{\Sigma}^{-1} E\left[\varepsilon_i \Big| \varepsilon_i \in (v_{i,j-1} - x_i'\beta,\ v_{i,j} - x_i'\beta], X = x_i\right]$$

This is the sum of Squared Generalized Residuals. The identity (3.5) implies that $\overline{S} = E(\overline{\varepsilon}'\overline{\varepsilon} \mid X)$ and $S = E(\varepsilon'\varepsilon \mid X)$ have the same derivatives with respect to $\beta$; consequently they are identical except for a constant. When we decompose the residual variance into the sum of between- and within-variance $E(\varepsilon'\varepsilon \mid X) = E(\overline{\varepsilon}'\,\overline{\varepsilon} \mid X) + E(\ddot{\varepsilon}\,'\,\ddot{\varepsilon} \mid X)$, it is obvious that this constant difference is just the within-variance $S - \overline{S} = E(\ddot{\varepsilon}\,'\,\ddot{\varepsilon} \mid X)$, which appears not to depend on $\beta$. Things are complicated since each individual $i$ has its own observation grid $\mathbb{C}_i$.

The solution for $\beta$ is found as the root-vector of the $K$-equation system $\overline{g}(\beta,\Sigma \mid X) = 0$.

We have now to construct the sample analogue of $\overline{g}(\beta,\Sigma \mid X)$. The $\overline{\varepsilon}_{i,j_i}$ have not drawn much attention in the empirical literature. One notable exception is Heckman (1976) who appears to be the first author in econometric literature to recognize the importance of this expected residual, later in the econometric literature sometimes called the Heckman-term (see also Van de Ven and Van Praag (1981)),. They have been called by Gouriéroux et al. (1987) the generalized residuals. They used them in the analysis of residuals. If the exact errors $\varepsilon$ are standard normally distributed, then we have for the coarsened errors the well-known formula

$$
\begin{aligned}
\overline{\varepsilon}_{i,j_i} &\overset{def}{=} E\left(\varepsilon_i \,\middle|\, \nu_{j_i-1} - x_i\beta < \varepsilon_i \le \nu_{j_i} - x_i\beta, X\right) = \\
&= \frac{\varphi(\nu_{j_i-1} - x_i\beta) - \varphi(\nu_{j_i} - x_i\beta)}{\Phi(\nu_{j_i} - x_i\beta) - \Phi(\nu_{j_i-1} - x_i\beta)}
\end{aligned}
\tag{3.6}
$$

---

[5] Notice that $\dfrac{\partial}{\partial\beta}\,\overline{\varepsilon} = \dfrac{\partial}{\partial\beta} E\left(\varepsilon \big| v_{j-1} - x\beta < \varepsilon \le v_j - x\beta, X = x\right) = \dfrac{\partial}{\partial\beta}\left[E\left(y \big| v_{j-1} < y \le v_j\right) - x\beta, X = x\right] = x$.

If the errors are not normally distributed but logistically, the formulas for the truncated marginal distribution can be found for example in Johnson, Kotz, Balakrishnan (1994) or in Maddala (1983, p.369). We shall restrict ourselves to the assumption of normally distributed errors.

Since there are no natural units observed, we can only estimate the $\beta$'s in (2.1) up to their ratios. A way to make them identifiable is to assume $\sigma_k = 1$ for $k = 1, \ldots, K$, which is the traditional assumption in Probit and item response analysis.

The sample moment analogue of (3.1) is

$$
\frac{1}{N}\sum_{i=1}^{N}\left[x_i\Sigma^{-1}\overline{e}_{i,j_i}\right] = \frac{1}{N}\sum_{i=1}^{N}\left[x_i\Sigma^{-1}\frac{\varphi(\nu_{j_i-1} - x_i\beta) - \varphi(\nu_{j_i} - x_i\beta)}{\Phi(\nu_{j_i} - x_i\beta) - \Phi(\nu_{j_i-1} - x_i\beta)}\right] = \hat{\overline{g}}(\beta) = 0
\tag{3.7}
$$

where $j_i$ is the index of the interval/block observed for the observation unit $i$.

Notice that (3.7) is a concise presentation of a system of $K$ blocks of $M$ equations, each corresponding with one of the elements of the coefficient matrix $\beta$, where we assume that each of the $K$ blocks contains $M$ different coefficients $\beta_k$.

The cut-points $\nu$ remain to be estimated. There are $K \times (J-1)$ of them. Therefore, we derive another additional set of $\nu$-identifying equations. The cut-points $\nu$ can be easily estimated one by one by applying the following strategy (called binarization). We define for each equation the $J-1$ auxiliary binary variables $\overline{\varepsilon}_{i,j}^{b}$ which may assume the lower value $E(\varepsilon_i|\varepsilon_i \leq v_j - x_i\beta)$ or the upper value $E(\varepsilon_i|\varepsilon_i > v_j - x_i\beta)$.[6] We have

$$P(\varepsilon_i \leq v_j - x_i\beta).E(\varepsilon_i|\varepsilon_i \leq v_j - x_i\beta) + P(\varepsilon_i > v_j - x_i\beta).E(\varepsilon_i|\varepsilon_i > v_j - x_i\beta) = 0 \qquad (3.8)$$

Again, there holds $E(\overline{\varepsilon}_{i,j}^{b}) = 0$, due to LIE. For the sample counterparts this implies

$$\text{plim}\left[\frac{1}{N}\sum \overline{e}_{i,j}^{b}\right] = 0, \ j = 1, 2, ..., J-1.$$

The sample moment analogues are

$$\frac{1}{N}\left[\sum_{i(j_i \leq j)} \frac{\varphi(v_{k,j} - x_{i,k}\beta_k)}{\Phi(v_{k,j} - x_{i,k}\beta)} - \sum_{i(j_i > j)} \frac{\varphi(v_{k,j} - x_{i,k}\beta)}{1 - \Phi(v_{k,j} - x_{i,k}\beta)}\right] = 0, \ k = 1, ..., K, j = 1, ..., J-1$$

$$(3.8a)$$

from which the cut-points $v_{k,j}$ can be easily estimated, as both sums at the left increase in $v_{k,j}$. We notice that these observations are not yet weighted by an error-covariance matrix.

Summing up we are ending with two equation systems (3.7) and (3.8) from which the parameters $\beta$ and $v$ are estimated. This can be done by applying the Generalized Method of Moments (GMM) (Hansen (1982)). We refer to well-known textbooks such as Cameron and Trivedi (2005), Greene (2018a), and Verbeek (2017) for elaborate descriptions. The software can be found, e.g., in Stata.[7] We use an iterative calculation scheme. Starting with assuming $\beta = 0$, a first-round yields $\beta^{(1)}, v^{(1)}$. Taking these values as starting point we repeat this iterative procedure until convergence, which is rather rapidly reached. The GMM-method provides us with an estimate of the covariance matrix of $\hat{\overline{\beta}}, \ \hat{\overline{v}}$ as well, using the well-known 'sandwich' formula.

---

[6] This trick, called "binarization" is suggested by, e.g., Chris Muris, 2017. "Estimation in the Fixed-Effects Ordered Logit Model," *The Review of Economics and Statistics*, vol. 99(3), pages 465-477, July. The term has been used before in computer science.

[7] We use a mixture of our own software in Fortran and in Stata. See our computer program online.

**4. Polychoric correlations and Coefficients of Determination.**

Suppose we have two test items $Y_1, Y_2$ $(K = 2)$ available by which we may, for example, examine an individual or test the effect of a specific therapy or a response on a satisfaction question. For simplicity we assume both items are yes/no questions. Then we are of course interested to know how correlated the two test items $Y_1, Y_2$ and therewith the responses on the two items are. The latent correlation between items is known in the psychometric literature as the polychoric correlation. The literature on polychoric correlations is massive. We refer out of the host of excellent contributions to the seminal Olsson (1979), and the more recent Liu, Li, Yu & Moustaki (2021), Moss and Grønneberg (2023) for more analysis. The problem is clearly how to estimate correlations between latent variables $Y_1, Y_2$, if we only have a $2 \times 2$ coincidence table at our disposition. We propose the following method.

The latent variables are modelled like (2.1). The latent model is

$$Y_{i,k} = X_{i,k,1}\beta_1 + \ldots + X_{i,k,M}\beta_M + \beta_0 + \varepsilon_{i,k} \quad i = 1,\ldots,N, \ k = 1,\ldots,K \quad (4.1)$$

where in this case $k = 1,2$. In this case we have

$$\text{Cov}(Y_1, Y_2) = \beta_1' E(X_1 X_2')\beta_2 + \sigma_{1,2}(\varepsilon) \quad (4.2)$$

and more generally for a $K \times K$ coincidence table we find the $K \times K$ -covariance matrix

$$\text{Cov}(Y) = B'\Sigma_{XX}B + \Sigma_{\varepsilon\varepsilon} \quad (4.3)$$

where $B'$ stands for the $K \times M$ matrix of structural effects and $\Sigma_{\varepsilon\varepsilon}$ for the latent error covariance matrix. Now we derive the polychoric correlation from $\text{Cov}(Y)$ in the usual way, that is, $\rho(Y_1 Y_2) = \text{Cov}(Y_1 Y_2)\big/\sqrt{\sigma(Y_1).\sigma(Y_2)}$. The covariance matrix $\Sigma_{YY} = \text{Cov}(Y)$ is estimated as

$$\hat{\Sigma}_{YY} = \hat{B}'\hat{\Sigma}_{XX}\hat{B} + \hat{\Sigma}_{\varepsilon\varepsilon} \quad (4.4)$$

where $\hat{B}$ is the estimated matrix of regression coefficients, $\hat{\Sigma}_{XX} = \frac{1}{N}\sum_{i=1}^{N} X_i X_i'$, and $\hat{\Sigma}_{\varepsilon\varepsilon}$ the estimated full error-covariance matrix, as estimated in the Appendix.

We notice that in the case that there are no structural effects found, i.e. $B = o$, we still may have non-zero polychoric correlations due to correlated errors. The corresponding correlations are found from the covariance matrix $\hat{\Sigma}_{YY}$ in the usual way.

The matrices $B, \Sigma_{XX}$ are already consistently estimated. The latent (full) error-covariance matrix $\Sigma_{\varepsilon\varepsilon}$ is yet unknown. In the Appendix we demonstrate how $\Sigma_{\varepsilon\varepsilon}$ is consistently estimated.

We note that this method does not assume normality of the random vectors $X$ or $\varepsilon$. We may also assume, for example, $\varepsilon$ to be logistic. In those cases the formula (3.3) for the generalized residual has to be replaced by the corresponding formula for the logistic, or in fact, any distribution, provided that the covariance matrix is finite.

In the second application below, where we are estimating satisfactions, we present the estimated $8 \times 8$ polyserial correlations between satisfactions as the off-diagonal elements in Table 5. For the first application in Section 5 we might estimate the polyserial correlations as well but given the panel nature of the data, it is not very interesting.

The relative explanatory power of the equation estimates depends on the question how volatile the outcomes are due to random errors. Consider (5.1). We have

$$\text{Var}(Y_1) = \beta_1' E(X_1 X_1') \beta_1 + \sigma_{1,1}(\varepsilon) \tag{4.5}$$

An attractive measure of fit, that is explanatory power is the traditional coefficient of determination

$$R^2 = \frac{\beta_1' E(X_1 X_1') \beta_1}{\beta_1' E(X_1 X_1') \beta_1 + \sigma_{1,1}(\varepsilon)} = 1 - \frac{1}{\beta_1' E(X_1 X_1') \beta_1 + \sigma_{1,1}(\varepsilon)} \tag{4.6}$$

The sample analogue for the first equation is

$$\hat{R}_k^2 = \frac{\dfrac{1}{N} \sum_{i=1}^{N} \left[ \sum_{m=1}^{M} \hat{\beta}_{k,m} x_{i,k,m} \right]^2}{\dfrac{1}{N} \sum_{i=1}^{N} \left[ \sum_{m=1}^{M} \hat{\beta}_{k,m} x_{i,k,m} \right]^2 + 1} \tag{4.7}$$

where $\sigma_{1,1}(\varepsilon) = 1$, as postulated in Section 3 This is the same magnitude as proposed by McKelvey and Zavoina (1975). We notice that these numbers may be interpreted as coefficients of determination of the regression equations (5.1) for $k = 1$, $k = 2, ..., k = 5$, respectively. Of course, the regression is not performed as $Y_{i,k}$ is not observable per individual. However, the regression correlation coefficient can be estimated by a detour using (5.6) or (5.7). We call these satisfaction correlation coefficients. They measure the part of the satisfaction variation, which can be structurally explained by observable traits $X$. If there are $K$ equations, we get $\hat{R}_1^2, ..., \hat{R}_K^2$.

## 5. Two empirical examples and one simulation experiment.

In order to evaluate our method empirically we considered two data sets, both part of a 2009-2013 panel data-sequence from the German SOEP-panel data set and a block of eight satisfaction questions in wave 2013 of the SOEP data. The first model in Section 6a is a set of five time-panel Ordered Probit equations where the errors are correlated. We call this estimation method Feasible Multivariate Ordered Probit (FMOP). It can be generalized to an arbitrary number of panel waves. The second data set consists of eight seemingly unrelated eight cross-section satisfaction questions, where errors are correlated. It is estimated in Section 6b. We call this a Seemingly Unrelated Ordered Probit model (SUOP). In addition we present estimations on a simulated data seton request of one the referees to this paper. Online we present the program code.

Given the results of the method it becomes possible to estimate the latent full covariance matrix $\Sigma$ as well. We defer the description of how to estimate the full latent covariance matrix to the Appendix and the online program code.

5a. *Employment status evaluation on a German five-year panel data set.*

Now we apply the FMOP method to a specific data set. We choose the employment situation of German workers, where we do not pretend to make a study of German employment but merely test the feasibility of the method, using these employment data. Following the lines above, we try to estimate the employment equation and the error covariance matrix using FMOP.

The data are derived from the German, Socio-Economic Panel (SOEP) data set. Households are followed for a period of five successive years (2009–2013). We assume an unstructured error covariance matrix. All explanatory variables are measured as deviations from their averages.

We use the variable employment (variable e11103 in the German SOEP data set) in three self-reported categories 'not working,' 'part-time working,' and 'full-time working.' This implies that the five grids for the years 2009–2013 consist of three intervals each.

We assume the explanatory variables 'age (18–75 years of age),' 'age-squared,' dummy variables for 'gender (female=1)', 'marital status: married (reference),' 'marital status: single,' 'marital status: separated,' 'ln(household income minus individual labour income),' 'number of children at home,' 'years of education,' and dummy variables for 'years of education unknown' and 'living in East-Germany.' The latent variable is assumed to be generated by the linear equation

$$\begin{aligned}\text{Employment} = {} & \beta_1\text{Age} + \beta_2\text{Age}^2 + \beta_3\text{Female} + \beta_4\text{Single} + \beta_5\text{Separated} + \\ & + \beta_6\text{Ln(HH.LabourInc)} + \beta_7\text{Children} + \beta_8\text{Years\_educ} + \\ & + \beta_9\text{Years\_educ\_unknown} + \beta_{10}\text{East} + \varepsilon\end{aligned} \quad (5.1)$$

As already said, we assume an 'unstructured' $5 \times 5$ covariance matrix where $\overline{\Sigma} = (\overline{\rho}_{k,k'})$. The results are presented in Table 1. In the left-hand panel we show the in-between error covariance matrix $\hat{\overline{\Sigma}}$, in the right-hand panel the latent full covariance matrix $\hat{\Sigma}$ as estimated by the simulation method described in the Appendix. The error-correlation over time appears from the right-panel to be quite considerable (1.0, .8775, .7800,…). When we look at the coarsened data the correlation is mitigated by the coarse observation but still considerable.

Table 1. In-between and full error covariance matrices for FMOP

|  | In-between covariance matrix (FMOP) | | | | | Full covariance matrix (FMOP) | | | | |
|---|---|---|---|---|---|---|---|---|---|---|
|  | 2009 | 2010 | 2011 | 2012 | 2013 | 2009 | 2010 | 2011 | 2012 | 2013 |
| 2009 | .6303 |  |  |  |  | 1.0000 |  |  |  |  |
| 2010 | .4760 | .6156 |  |  |  | .8775 | 1.0000 |  |  |  |
| 2011 | .4049 | .4746 | .6149 |  |  | .7800 | .8856 | 1.0000 |  |  |
| 2012 | .3695 | .4125 | .4690 | .6131 |  | .7343 | .7998 | .8837 | 1.0000 |  |
| 2013 | .3336 | .3642 | .4059 | .4833 | .6168 | .6867 | .7280 | .7950 | .9079 | 1.0000 |

The regression estimates according to FMOP and Ordered Probit (errors independent) are presented in Table 2.

Table 2. Regression estimates from FMOP and Ordered Probit.

| | Feasible Multivariate Ordered Probit (FMOP) | | | Exact observ. | Ordered Probit (OP) | | |
|---|---|---|---|---|---|---|---|
| | Number of obs.=51760 | | | FMOP | Number of obs.=51760 McF pseudo R²=0.2523 McK-Z R²=0.5396 | | |
| EMPLOYMENT | Coeff. | Std.err. | z | Std.err. | Coeff. | Std.err. | z |
| AGE | .3137 | .0059 | 53.50 | .0038 | .3236 | .0032 | 102.36 |
| AGE square | -.0037 | .00006 | -58.34 | .00004 | -.0038 | .00003 | -114.77 |
| D. FEMALE | -.7464 | .0223 | -33.51 | .0180 | -.7727 | .0115 | -67.01 |
| D. SINGLE | .0778 | .0309 | 2.52 | .0186 | .0541 | .0202 | 2.68 |
| D. SEPARATED | .0538 | .0250 | 2.15 | .0164 | .1184 | .0175 | 6.77 |
| Ln(HHLABOURINC) | .0071 | .0015 | 4.65 | .0009 | .0150 | .0013 | 11.31 |
| # of CHILDREN | -.2051 | .0114 | -17.97 | .0074 | -.2571 | .0077 | -33.47 |
| YEARS EDUCATION | .0789 | .0039 | 20.03 | .0031 | .0742 | .0021 | 34.97 |
| D. EDUC. unknown | -.1172 | .0459 | -2.55 | .0318 | -.1985 | .0294 | -6.75 |
| D. EAST Germany | -.0542 | .0237 | -2.28 | .0180 | -.0420 | .0128 | -3.28 |
| Cut point 1 | -.5892 | .0129 | -45.79 | -- | -.5984 | .0069 | -86.68 |
| Cut point 2 | .3569 | .0123 | 29.05 | -- | .3753 | .0066 | 5.68 |

As expected the regression estimates are of the same order, because both estimators are consistent. The difference is clearly in the calculated standard deviations. All FMOP standard errors are a factor 1,5 to 2,0 larger than the OP-estimates. This is caused because the assumption of error independence by OP instead of the observed strong error correlations is tantamount to a gross exaggeration of the reliability of the data material when we ignore the non-zero error correlations. The difference in standard deviations is a warning signal.

For curiosity we also look at the question what standard deviations we would have found when we would have had the non-coarsened, that is exact, data material at our disposal. Those standard deviations are estimated by the roots of the diagonal elements of $\frac{1}{N}(X'\hat{\Sigma}^{-1}X)^{-1}$ the elements of which are known. The latent error-covariance matrix $\Sigma$ is estimated by $\hat{\Sigma}$ according to the method described in the Appendix. We see from comparison that the FMOP-standard deviations (e.g. for the AGE-coefficient 0.3137) on the basis of the coarsened observations are much larger than the corresponding theoretical values (0.0038) for GLS-estimation on the exact latent data.

The computation time in total was 8 seconds. We used a laptop. The computation process can be split up into two parts: the first-stage OP estimation, taking 3 seconds and the second-stage estimation taking another 5 seconds.

We see that employment increases with age until age 42, after which employment decreases (we excluded respondents under-18s and the over-75s). Females are less often employed than males. In households with children the respondents work less than in childless households. The more additional labour income in the household, the more the respondent works. The more education years one has, the more one works full-time, while respondents from East Germany are less employed than the West Germans.

5b. *Seemingly Unrelated Ordered Probit (SUOP) on a block of eight satisfaction questions.*

In the German panel questionnaire, we find a number of satisfaction questions referring to various life domains, like those presented in Fig. 1. Here, we apply the SUOP method.

This type of questioning is abundantly used in marketing research and happiness research. Another very important instance, where the use of SUOP is at hand, is in the analysis of vignettes, also known as factorial surveys in sociological research or as conjoint analysis, now one of the major tools in psychology and marketing research (Green and Srinivasan (1978), Atzmüller and Steiner (2010), Wallander (2009), Van Beek et al. (1997)).

| How satisfied are you today with the following areas of your life? Please answer on a scale from 0 to 10, where 0 means completely dissatisfied and 10 means completely satisfied. How satisfied are you with ..... | |
| --- | --- |
| 1. your health? | 0 1 2 3 4 5 6 7 8 9 10 |
| 2. your sleep? | 0 1 2 3 4 5 6 7 8 9 10 |
| 3. your household income? | 0 1 2 3 4 5 6 7 8 9 10 |
| 4. your personal income? | 0 1 2 3 4 5 6 7 8 9 10 |
| 5. your dwelling? | 0 1 2 3 4 5 6 7 8 9 10 |
| 6. your leisure time? | 0 1 2 3 4 5 6 7 8 9 10 |
| 7. your family life? | 0 1 2 3 4 5 6 7 8 9 10 |
| 8. your standard of living? | 0 1 2 3 4 5 6 7 8 9 10 |

Fig. 1. A block of satisfaction questions with respect to various life domains.

The data set consists of about 15,000 observation units. Since the original formulation with 11 answer categories made the coarsened observations look very similar to continuous observations, we further coarsened the data into five response categories (0,1,2), (3,4),...,(9,10). In this paper we apply SUOP analysis to the above listed block of satisfaction questions with respect to life domains from the 2013 wave of the GSOEP panel. We use the following explanatory variables: age and age-squared, dummies for being female, single and separated, ln(individual labour income), ln(household income *minus* individual labour income), the number of children, the number of years of education, living in East Germany, dummy disability status (0 (no), 1 (yes)), and health rating (1 (bad health),…,5 (very good health)). Our primary objective is to demonstrate the feasibility of SUOP. It stands to reason that for a substantive analysis of domain satisfactions this model specification is probably too simplistic, however, for our objective, testing the feasibility of SUOP, this specific choice is no problem. In order to avoid that every dependent variable would be explained by the same set of explanatory variables we chose different subsets for each equation.

In Table 3 we present the estimates of the first two equations on Health and Sleep satisfaction. For the full table presenting all eight equation estimates we refer to the Appendix.

In the first two columns we present the initial Probit estimates and their s.e.'s. In columns 3, 4 we present the corresponding SUOP-estimates and their s.e's. The two right-hand columns 5, 6 give the corresponding estimates by means of the ML-method. We take the `cmp` results as the touchstone of our comparison.

Table 3. Comparison of the parameter estimates and their s.e.'s for Ordered Probit, Method of Moments, and Maximum Likelihood.

| McK-Z $R^2$ | 0.1489 | | 0.1481 | | 0.1154 | |
|---|---|---|---|---|---|---|
| *Health Satisfaction* | $\beta_{OP}$ | $\sigma_{OP}$ | $\beta_{SUOP}$ | $\sigma_{SUOP}$ | $\beta_{ML}$ | $\sigma_{ML}$ |
| Health: AGE | -.0695 | .0046 | -.0693 | .0048 | -.0651 | .0045 |
| Health: AGE$^2$ | .0006 | .00005 | .0006 | .00005 | .0006 | .00005 |
| Health: D_FEMALE | -.0182 | .0175 | -.0181 | .0174 | -.0134 | .0166 |
| Health: D_SINGLE | -.0514 | .0291 | -.0510 | .0293 | -.0570 | .0278 |
| Health: D_SEPARATED | -.1241 | .0240 | -.1238 | .0243 | -.1108 | .0230 |
| Health: Ln_LABOURINC | .0289 | .0025 | .0288 | .0026 | .0227 | .0022 |
| Health: CHILDREN | .0357 | .0123 | .0356 | .0126 | .0308 | .0114 |
| Health: D_EAST | -.1678 | .0201 | -.1670 | .0196 | -.1402 | .0193 |
| Health: DISABLE | -.7991 | .0272 | -.7971 | .0268 | -.5825 | .0232 |
| Health: Cut point 1 | -1.8277 | .0184 | -1.8240 | .0191 | -1.8163 | .0181 |
| Health: Cut point 2 | -1.1161 | .0131 | -1.1094 | .0134 | -1.1122 | .0130 |
| Health: Cut point 3 | -.3432 | .0109 | -.3415 | .0109 | -.3399 | .0107 |
| Health: Cut point 4 | .9468 | .0123 | .9442 | .0124 | .9409 | .0122 |
| McK-Z $R^2$ | 0.0278 | | 0.0278 | | 0.0196 | |
| *Sleep Satisfaction* | $\beta_{OP}$ | $\sigma_{OP}$ | $\beta_{SUOP}$ | $\sigma_{SUOP}$ | $\beta_{ML}$ | $\sigma_{ML}$ |
| Sleep: AGE | -.0380 | .0041 | -.0381 | .0042 | -.0348 | .0041 |
| Sleep: AGE$^2$ | .0004 | .00004 | .0004 | .00004 | .0003 | .00004 |
| Sleep: D. FEMALE | -.1213 | .0172 | -.1213 | .0172 | -.1204 | .0164 |
| Sleep: D. SINGLE | .0508 | .0302 | .0505 | .0307 | .0594 | .0284 |
| Sleep: D. SEPARATED | -.0762 | .0253 | -.0762 | .0260 | -.0658 | .0238 |
| Sleep: Ln(HH.LABOURINC) | .0067 | .0021 | .0066 | .0020 | .0062 | .0018 |
| Sleep: # of CHILDREN | .0219 | .0121 | .0220 | .0122 | .0164 | .0113 |
| Sleep: YEARS of EDUCATION | .0316 | .0032 | .0315 | .0031 | .0117 | .0028 |
| Sleep: D. EAST GERMANY | -.1063 | .0199 | -.1065 | .0196 | -.0763 | .0192 |
| Sleep: Cut point 1 | -1.6918 | .0175 | -1.6912 | .0174 | -1.7036 | .0174 |
| Sleep: Cut point 2 | -.9791 | .0123 | -.9800 | .0123 | -.9831 | .0123 |
| Sleep: Cut point 3 | -.9251 | .0106 | -.2970 | .0106 | -.2940 | .0104 |
| Sleep: Cut point 4 | .7286 | .0114 | .7269 | .0115 | .7150 | .0113 |

Our first conclusion is that the three methods OP, SUOP, ML yield estimates which do not differ significantly in most cases. This is not surprising as the three estimators are consistent. The standard deviations of the SUOP-estimators seem to be slightly larger than the ML-estimators, but the differences are mostly negligible.

In Table 4 we present the full correlation matrices as estimated by SUOP (estimation according to Appendix) and ML (according to Stata), respectively.

Table 4. Full error correlation matrices compared for SUOP and ML.

| Residual Corr. SUOP | Health | Sleep | HH inc. | Ind inc. | Dwelling | Leisure | Family life | Stand. living |
|---|---|---|---|---|---|---|---|---|
| Health | 1.0000 | | | | | | | |
| Sleep | .5328 | 1.0000 | | | | | | |
| HH. inc. | .3196 | .2758 | 1.0000 | | | | | |
| Ind. inc. | .2851 | .2367 | .8414 | 1.0000 | | | | |
| Dwelling | .3103 | .2894 | .4521 | .3769 | 1.0000 | | | |
| Leisure | .3246 | .3359 | .3274 | .2812 | .4557 | 1.0000 | | |
| Family life | .3585 | .3295 | .3327 | .2838 | .4526 | .4743 | 1.0000 | |
| Stand. living | .3964 | .2520 | .7173 | .6044 | .5551 | .4626 | .5930 | 1.0000 |
| Residual Corr. ML | Health | Sleep | HH inc. | Ind inc. | Dwelling | Leisure | Family life | Stand. living |
| Health | 1.0000 | | | | | | | |
| Sleep | .5487 | 1.0000 | | | | | | |
| HH. inc. | .3280 | .2802 | 1.0000 | | | | | |
| Ind. inc. | .2919 | .2487 | .8266 | 1.0000 | | | | |
| Dwelling | .3144 | .2990 | .4530 | .3869 | 1.0000 | | | |
| Leisure | .3301 | .3463 | .3323 | .2899 | .4635 | 1.0000 | | |
| Family life | .3643 | .3405 | .3355 | .2849 | .4563 | .4800 | 1.0000 | |
| Stand. living | .3973 | .3492 | .7099 | .6118 | .5582 | .4610 | .5930 | 1.0000 |

We see that of the 75 SUOP-estimated regression coefficients 17 fall out of the ML-confidence intervals. For the estimates of the correlation matrix we find a similar result.[8] Three of the 28 SUOP estimated correlation coefficients are just outside the ML-confidence intervals.

---

[8] The cmp-procedure provides confidence intervals for the correlation estimates.

The polychoric correlation matrix is presented in Table 5a.

Table 5a. The polychoric correlation matrix.

| Residual Corr. SUOP | Health | Sleep | HH inc. | Ind inc. | Dwelling | Leisure | Family life | Stand. living |
|---|---|---|---|---|---|---|---|---|
| Health | 1.0000 | | | | | | | |
| Sleep | .7372 | 1.0000 | | | | | | |
| HH. inc. | .7626 | .6347 | 1.0000 | | | | | |
| Ind. inc. | .7389 | .6028 | .9418 | 1.0000 | | | | |
| Dwelling | .6985 | .5998 | .7752 | .7338 | 1.0000 | | | |
| Leisure | .6962 | .6258 | .7111 | .6704 | .7287 | 1.0000 | | |
| Family life | .7496 | .6372 | .7550 | .7101 | .7558 | .7521 | 1.0000 | |
| Stand. living | .8015 | .6728 | .9220 | .8705 | .8169 | .7680 | .8538 | 1.0000 |

A naïve approach is to assign the values 0,1,…,9,10 to the satisfaction values and to calculate the Pearson correlations on that basis. This assignment is conform to daily usage, where average satisfaction values in a sample are also based on this assignment practice. The Pearson correlations are presented in Table 5b. We see that there is a considerable difference between Table 5a and 5b we prefer 5a to 5b, as the cardinalization by 0,1..,10 is arbitrary and might be replaced by another one (0,2,3,…) yielding a different Table 5b, while the polyserial correlations are based on an endogenous cardinalization. We notice that all Pearson correlations in Table 5b are considerably smaller than the corresponding numbers in Table 5a.

Table 5b. The Pearson correlation matrix. (scale 0-10)

| Pearsoncorr. scale (0-10) | Health | Sleep | HH inc. | Ind inc. | Dwelling | Leisure | Family life | Stand. living |
|---|---|---|---|---|---|---|---|---|
| Health | 1.0000 | | | | | | | |
| Sleep | .5185 | 1.0000 | | | | | | |
| HH. inc. | .3331 | .2829 | 1.0000 | | | | | |
| Ind. inc. | .2899 | .2575 | .7761 | 1.0000 | | | | |
| Dwelling | .2414 | .2635 | .4307 | .3580 | 1.0000 | | | |
| Leisure | .2121 | .2804 | .2921 | .2448 | .4047 | 1.0000 | | |
| Family life | .3027 | .2965 | .3331 | .2599 | .4085 | .3985 | 1.0000 | |
| Stand. living | .3852 | .3325 | .7112 | .5872 | .5201 | .3806 | .5387 | 1.0000 |

The computation process can be split up into three parts: the first-stage OP estimation took 1.8 seconds on our ASUS VivoBook 15 laptop, and the second stage SUOP-estimation took another 111 seconds. The whole calculation requires less than two minutes. The ML-estimations by `cmp` (default) took 7 hours. In the SUOP-method we use the in-between covariance matrix $\hat{\hat{\Sigma}}$ and not the full covariance estimate. The computation time depends on the sample size $N$, the size $K$ of the error covariance matrix, and the capacity of our laptop. In this example $K$ =8. We see that for the ML-method the time increases non-linearly with $N$. The number $K$ seems to be important as well. For $K$ =2 equations both methods are roughly equally fast, where SUOP takes 11 seconds and ML-cmp 10.5 seconds for $N$ =15535. For $K$ =3 the SUOP computation increases to 16 seconds, while the ML-method requires already 1,241 seconds. This is caused, it seems, by the fact that ML has to evaluate a lot of $K$-dimensional integrals. A colleague of ours (an expert Stata-user) observed, quoting the 'options' in the Stata text, that `cmp` uses the GHK-simulation method for evaluating the needed integrals and that in the default option cmp uses $2\sqrt{N}$ draws per evaluated likelihood. In the present case this is about 250 simulations per observation. Capellari and Jenkins (2003) suggested that for a large number of observations the number of draws can be considerably reduced without severe efficiency loss. According to our colleague by taking 5 draws per likelihood we would reduce the computation time from the reported 7 hours to 8 minutes with only a slight efficiency reduction. That is probably still significantly slower than the new method, but the revision would be material. We followed this suggestion and found indeed comparable estimates for the coefficients $\beta$. To our surprise the standard deviations for the five draws were not significantly different from the 250 draws version. This seems to indicate that in the assessment of variance the additional contribution caused by the simulation variance is not taken into account.

Clearly, if we would reduce the number of equations from eight to a more manageable four or two equations and/or reduce the number of observations, both the ML and the SUOP-methods would perform much faster.

Our conclusion is that the SUOP-method is faster than the ML-method. We are unable to say whether the Stata procedure cmp is to blame and could be improved or whether this is a general feature of the ML-GHK procedure. It might also be that we could have reduced the ML-computation time by choosing specific options instead of the default procedure. Choosing too severe tolerance levels for the iterations involved would have increased the computation time in exchange for more exact confidence intervals. However, given that the `cmp`-outcomes have about the same confidence intervals as our SUOP outcomes we do not believe that the tolerance levels chosen in `cmp` were more severe than in our method.

## 5c. *A simulated example.*

Finally, we apply the estimation method to a simulated data set. We simulated a hard data set of 10,000 observations. We generated the set as follows. We assumed a latent model

$$
\begin{aligned}
y_{i,1} &= \ x_{i,1} + \quad\ x_{i,2} + \ x_{i,3} + \ x_{i,4} + \varepsilon_{i,1} \\
y_{i,2} &= \ x_{i,1} + \ 2x_{i,2} + 3x_{i,3} + 4x_{i,4} + \varepsilon_{i,2} \\
y_{i,3} &= -x_{i,1} + \ 2x_{i,2} - 3x_{i,3} + 4x_{i,4} + \varepsilon_{i,3} \\
y_{i,4} &= -x_{i,1} + 0.5x_{i,2} - \ x_{i,3} + \ x_{i,4} + \varepsilon_{i,4}
\end{aligned}
$$

where, $x_1$ is normal $N(0,1)$, where $x_2 = 0.5x_1 + D1$ with $D1$ a dummy variable equal to +1 or -1 with 50% each, where $x_3 = N(0,2) + 0.5x_2$, and where $x_4 = -0.5x_3 + D2$ and $D2$ is drawn to equal +1, 0, or -1 with a chance of $\tfrac{1}{3}$ each.

The error vector $\varepsilon$ is i.i.d. $N(0,\Sigma)$ with

$$
\Sigma = \begin{pmatrix}
1.0 & & & \\
0.5 & 1.0 & & \\
-0.5 & 0.3 & 1.0 & \\
0.2 & 0.6 & -0.1 & 1.0
\end{pmatrix}.
$$

We notice that all four variables $x$ and the error vector have expectation equal to zero. In order to avoid that (the non-conditioned) $Y_i$ is approximately normal, we restricted the explanatory variables to a small number of four and we chose those variables to be non-normal and correlated, such that the structural part $\beta'x$ does not tend to normality. Our first aim is to look for the distribution of the exact data. The expectation $E(Y) = 0$, the empirical mean equals 0.0160 and the variance var($Y$) equals 2.603. The correlation matrix of the variables $X$, and $Y$ is the 8×8 matrix in Table 1.

Table 6c.1. The correlation matrix of the variables ($N$=10,000)

|    | x1     | x2     | x3     | x4     | y1     | y2     | y3     | y4     |
|----|--------|--------|--------|--------|--------|--------|--------|--------|
| x1 | 1.0000 |        |        |        |        |        |        |        |
| x2 | .8797  | 1.0000 |        |        |        |        |        |        |
| x3 | -.3812 | -.2516 | 1.0000 |        |        |        |        |        |
| x4 | -.4803 | -.3452 | .9240  | 1.0000 |        |        |        |        |
| y1 | .6278  | .4450  | -.1164 | -.3281 | 1.0000 |        |        |        |
| y2 | .7033  | .6220  | -.0943 | -.2136 | .4566  | 1.0000 |        |        |
| y3 | .5598  | .5484  | -.9312 | -.8929 | .1250  | .2659  | 1.0000 |        |
| y4 | -.2500 | -.0479 | .9196  | .8447  | -.0921 | -.2116 | -.7871 | 1.0000 |

Cut points are defined as

$$\nu_{1,1} = \;\; 0$$
$$\nu_{2,1} = -1, \quad \nu_{2,2} = \;\; 1.5$$
$$\nu_{3,1} = -1, \quad \nu_{3,2} = \;\; 0.5$$
$$\nu_{4,1} = -1.5, \;\; \nu_{4,2} = -0.5, \;\; \nu_{4,3} = 1$$

We define the response indicators:

$$j_{i,1} = 1,2$$
$$j_{i,2} = 1,2,3$$
$$j_{i,3} = 1,2,3$$
$$j_{i,4} = 1,2,3,4$$

The model is iteratively estimated by the FMOP-method. $j_{i,k}$ is the interval index by respondent $i$ for equation $k$, corresponding with the four equations $k$=1,...,4. We start with iteration $t$=0 for $\beta = 0$. We define the under- and upper residuals

$$E(\varepsilon \,|\, \varepsilon \leq \nu_{j_{i,k}}^{(t)} - \beta_k^{(t)'} x_{i,k}) = \overline{e}_{j_{i,k}}^{(t)} = \frac{-\varphi(\nu_{j_{i,k}}^{(t)} - \beta_k^{(t)'} x_{i,k})}{\Phi(\nu_{j_{i,k}}^{(t)} - \beta_k^{(t)'} x_{i,k})}$$

$$E(\varepsilon \,|\, \varepsilon > \nu_{j_{i,k}}^{(t)} - \beta_k^{(t)'} x_{i,k}) = \tilde{e}_{j_{i,k}}^{(t)} = \frac{\varphi(\nu_{j_{i,k}}^{(t)} - \beta_k^{(t)'} x_{i,k})}{1 - \Phi(\nu_{j_{i,k}}^{(t)} - \beta_k^{(t)'} x_{i,k})}$$

We define the sets of respondents $S_{k,j}^1$, $S_{k,j}^2$ ($k$=1,...4; $j$=1,...,$J_k$) who are in the response categories $\leq j$ or $> j$ respectively. We solve the equations

$$\sum_{i \in S_{k,j}^1} \overline{e}_{j_{i,k}}^{(t)} + \sum_{i \in S_{k,j}^2} \tilde{e}_{j_{i,k}}^{(t)} = 0 \,, \; k = 1,...,4, \, j = 1,...,J_k \qquad (5c.1)$$

for $v_{k,j}^{(t)}$ and find estimated cut-points $v_{k,j}^{(t)}$ in the $t^{\text{th}}$ iteration. These cut-points $v_{k,j}^{(t)}$ are substituted to define the generalized residuals. The estimated generalized residuals $\overline{e}_{j_{i,k}}^{(t)}$ in the $t^{th}$ iteration are

$$\overline{e}_{j_{i,k}}^{(t)} = \frac{\varphi(v_{j_{i,k}-1}^{(t)} - \beta_k'x_{i,k}) - \varphi(v_{j_{i,k}}^{(t)} - \beta_k'x_{i,k})}{\Phi(v_{j_{i,k}-1}^{(t)} - \beta_k'x_{i,k}) - \Phi(v_{j_{i,k}-1}^{(t)} - \beta_k'x_{i,k})},$$

where $j_{i,k}$ is the interval index by respondent $i$ for equation $k$. corresponding with the four equations $k$=1,...,4. We now define the $(K \times M)$ orthogonality conditions

$$\frac{1}{N}\sum_{i=1}^{N} x_{i,k,m}\overline{e}_{j_{i,k}}^{(t)} = 0 , \ k = 1,...,4, \ m = 1,...,M \qquad (5c.2)$$

We have now two equation systems (5c.1) and (5c.2), which are simultaneously solved. Then we calculate the in-between error covariance matrix

$$\overline{\Sigma}^{(t)} = \left(\overline{\sigma}_{k,k'}\right) = \frac{1}{N}\left(\sum_{i=1}^{N}\overline{e}_{j_{i,k}}^{(t)}\overline{e}_{j_{i,k'}}^{(t)}\right).$$

We repeat (5c.1) and (5c.2) with the new $\beta^{(t+1)}$ and $v_{k,j}^{(t+1)}$, and find new estimates. We repeat (5c.1) and (5c.2) after weighting with the inverse covariance matrix solving

$$\sum_{i=1}^{N}\overline{e}_{j_{i,k}}^{(t)}\left[\overline{\Sigma}^{(t)}\right]^{-1} x_{i,k} = 0 .$$

In the end we estimate the corresponding covariance matrix of the estimators $\hat{\overline{\beta}}, \ \hat{\overline{v}}$ by the well-known sandwich formula.

We estimate each non-diagonal element $\sigma_{k,k'}$ of the latent full covariance matrix from the corresponding element $\overline{\sigma}_{k,k'}$ of the in-between error covariance matrix. The method is described in detail in the Appendix.

Table 5c.2a. Beta's, Standard errors and Error correlations (*N*=10,000)
(Gray means, the coefficient is NOT in the 95% confidence interval of cmp)

| N=10000 | beta's MoM | s.e. MoM | beta's cmp | s.e. cmp |
|---|---|---|---|---|
| Eq.y1: var. x1 | 0.9794 | 0.0264 | 0.9799 | 0.0269 |
| Eq.y1: var. x2 | 0.9910 | 0.0240 | 0.9927 | 0.0241 |
| Eq.y1: var. x3 | 0.9715 | 0.0222 | 0.9715 | 0.0223 |
| Eq.y1: var. x4 | 0.9694 | 0.0281 | 0.9703 | 0.0282 |
| Eq.y1: cutp. 1 | -0.0024 | 0.0193 | -0.0072 | 0.0191 |
| Eq.y2: var. x1 | 1.0329 | 0.0290 | 1.0557 | 0.0279 |
| Eq.y2: var. x2 | 1.9859 | 0.0396 | 1.9991 | 0.0387 |
| Eq.y2: var. x3 | 3.0286 | 0.0550 | 3.0558 | 0.0535 |
| Eq.y2: var. x4 | 4.0158 | 0.0733 | 4.0526 | 0.0711 |
| Eq.y2: cutp. 1 | -1.0215 | 0.0321 | -1.0154 | 0.0311 |
| Eq.y2: cutp. 2 | 1.4972 | 0.0386 | 1.5026 | 0.0358 |
| Eq.y3: var. x1 | -1.0234 | 0.0444 | -1.0065 | 0.0406 |
| Eq.y3: var. x2 | 2.0183 | 0.0640 | 1.9981 | 0.0586 |
| Eq.y3: var. x3 | -3.0813 | 0.0847 | -3.0441 | 0.0805 |
| Eq.y3: var. x4 | 4.1387 | 0.1152 | 4.0610 | 0.1097 |
| Eq.y3: cutp. 1 | -0.9753 | 0.0447 | -0.9675 | 0.0428 |
| Eq.y3: cutp. 2 | 0.5115 | 0.0407 | 0.4993 | 0.0387 |
| Eq.y4: var. x1 | -1.0079 | 0.0212 | -1.0078 | 0.0209 |
| Eq.y4: var. x2 | 0.4957 | 0.0169 | 0.4931 | 0.0165 |
| Eq.y4: var. x3 | -1.0180 | 0.0173 | -1.0143 | 0.0171 |
| Eq.y4: var. x4 | 0.9727 | 0.0224 | 0.9696 | 0.0218 |
| Eq.y4: cutp. 1 | -1.5233 | 0.0282 | -1.5198 | 0.0273 |
| Eq.y4: cutp. 2 | -0.5125 | 0.0233 | -0.5165 | 0.0223 |
| Eq.y4: cutp. 3 | 1.0335 | 0.0253 | 1.0289 | 0.0245 |

| Estimated Full correlation matrix MoM | | | | Estimated Full correlation matrix cmp | | | |
|---|---|---|---|---|---|---|---|
| 1.0000 | | | | 1.0000 | | | |
| 0.4496 | 1.0000 | | | 0.4703 | 1.0000 | | |
| -0.5450 | 0.2527 | 1.0000 | | -0.4841 | 0.2306 | 1.0000 | |
| 0.1934 | 0.5877 | -0.0846 | 1.0000 | 0.2110 | 0.6266 | -0.0857 | 1.0000 |

Table 5c.2b. Beta's, Standard errors and Error correlations (*N*=5,000)
(Base dataset of 10000 case, only every second case is used)

| N=5000 | beta's MoM | s.e. MoM | beta's cmp | s.e. cmp |
|---|---|---|---|---|
| Eq.y1: var. x1 | 1.0192 | 0.0385 | 1.0161 | 0.0392 |
| Eq.y1: var. x2 | 0.9952 | 0.0341 | 0.9949 | 0.0344 |
| Eq.y1: var. x3 | 0.9598 | 0.0319 | 0.9573 | 0.0312 |
| Eq.y1: var. x4 | 0.9484 | 0.0398 | 0.9460 | 0.0397 |
| Eq.y1: cutp. 1 | -0.0330 | 0.0273 | -0.0361 | 0.0270 |
| Eq.y2: var. x1 | 1.0832 | 0.0418 | 1.0969 | 0.0402 |
| Eq.y2: var. x2 | 1.9317 | 0.0552 | 1.9383 | 0.0534 |
| Eq.y2: var. x3 | 3.0125 | 0.0790 | 3.0381 | 0.0749 |
| Eq.y2: var. x4 | 4.0011 | 0.1053 | 4.0333 | 0.0997 |
| Eq.y2: cutp. 1 | -1.0462 | 0.0462 | -1.0508 | 0.0442 |
| Eq.y2: cutp. 2 | 1.4679 | 0.0542 | 1.4743 | 0.0501 |
| Eq.y3: var. x1 | -1.0418 | 0.0636 | -1.0452 | 0.0567 |
| Eq.y3: var. x2 | 1.9988 | 0.0847 | 2.0070 | 0.0807 |
| Eq.y3: var. x3 | -3.0718 | 0.1113 | -3.0620 | 0.1128 |
| Eq.y3: var. x4 | 4.1625 | 0.1512 | 4.1297 | 0.1548 |
| Eq.y3: cutp. 1 | -0.9182 | 0.0619 | -0.8988 | 0.0598 |
| Eq.y3: cutp. 2 | 0.6152 | 0.0562 | 0.6031 | 0.0554 |
| Eq.y4: var. x1 | -0.9814 | 0.0300 | -0.9829 | 0.0292 |
| Eq.y4: var. x2 | 0.4882 | 0.0238 | 0.4860 | 0.0233 |
| Eq.y4: var. x3 | -1.0281 | 0.0246 | -1.0261 | 0.0244 |
| Eq.y4: var. x4 | 0.9506 | 0.0322 | 0.9492 | 0.0306 |
| Eq.y4: cutp. 1 | -1.5350 | 0.0398 | -1.5408 | 0.0390 |
| Eq.y4: cutp. 2 | -0.5381 | 0.0327 | -0.5369 | 0.0321 |
| Eq.y4: cutp. 3 | 1.0050 | 0.0362 | 1.0132 | 0.0343 |

| Estimated Full correlation matrix MoM | | | | Estimated Full correlation matrix cmp | | | |
|---|---|---|---|---|---|---|---|
| 1.0000 | | | | 1.0000 | | | |
| 0.5288 | 1.0000 | | | 0.4930 | 1.0000 | | |
| -0.3185 | 0.3089 | 1.0000 | | -0.3905 | 0.3245 | 1.0000 | |
| 0.3014 | 0.6468 | -0.1165 | 1.0000 | 0.2201 | 0.6138 | -0.0981 | 1.0000 |

0.3014: 95% confidence interval cmp [0.1462 : 0.2916]

Table5c.2c. Beta's, Standard errors and Error correlations (*N*=2,000)
(Base dataset of 10000 case, only every fifth case is used)

| N=2000 | beta's MoM | s.e. MoM | beta's cmp | s.e. cmp |
|---|---|---|---|---|
| Eq.y1: var. x1 | 0.9929 | 0.0574 | 0.9814 | 0.0590 |
| Eq.y1: var. x2 | 0.9259 | 0.0507 | 0.9186 | 0.0509 |
| Eq.y1: var. x3 | 0.9448 | 0.0512 | 0.9392 | 0.0483 |
| Eq.y1: var. x4 | 0.9605 | 0.0628 | 0.9593 | 0.0609 |
| Eq.y1: cutp. 1 | -0.0165 | 0.0418 | -0.0081 | 0.0414 |
| Eq.y2: var. x1 | 0.9729 | 0.0643 | 0.9747 | 0.0588 |
| Eq.y2: var. x2 | 1.9411 | 0.0822 | 1.9198 | 0.0827 |
| Eq.y2: var. x3 | 2.9799 | 0.1168 | 2.9654 | 0.1138 |
| Eq.y2: var. x4 | 3.9235 | 0.1548 | 3.9077 | 0.1502 |
| Eq.y2: cutp. 1 | -1.0068 | 0.0705 | -0.9718 | 0.0677 |
| Eq.y2: cutp. 2 | 1.4167 | 0.0828 | 1.4069 | 0.0746 |
| Eq.y3: var. x1 | -1.0846 | 0.0992 | -1.0725 | 0.0990 |
| Eq.y3: var. x2 | 2.1905 | 0.1320 | 2.2170 | 0.1474 |
| Eq.y3: var. x3 | -3.3765 | 0.1985 | -3.4092 | 0.2109 |
| Eq.y3: var. x4 | 4.4998 | 0.2566 | 4.5139 | 0.2813 |
| Eq.y3: cutp. 1 | -1.2270 | 0.1100 | -1.2034 | 0.1120 |
| Eq.y3: cutp. 2 | 0.5252 | 0.0976 | 0.5239 | 0.0939 |
| Eq.y4: var. x1 | -0.9667 | 0.0465 | -0.9641 | 0.0458 |
| Eq.y4: var. x2 | 0.4765 | 0.0372 | 0.4680 | 0.0362 |
| Eq.y4: var. x3 | -1.0109 | 0.0378 | -0.9994 | 0.0386 |
| Eq.y4: var. x4 | 0.9413 | 0.0511 | 0.9278 | 0.0472 |
| Eq.y4: cutp. 1 | -1.5391 | 0.0629 | -1.5537 | 0.0605 |
| Eq.y4: cutp. 2 | -0.5099 | 0.0509 | -0.5206 | 0.0493 |
| Eq.y4: cutp. 3 | 0.9616 | 0.0541 | 0.9675 | 0.0527 |

| Estimated Full correlation matrix MoM | | | | Estimated Full correlation matrix cmp | | | |
|---|---|---|---|---|---|---|---|
| 1.0000 | | | | 1.0000 | | | |
| 0.5982 | 1.0000 | | | 0.5334 | 1.0000 | | |
| -0.2635 | 0.3972 | 1.0000 | | -0.3878 | 0.2402 | 1.0000 | |
| 0.3114 | 0.6242 | 0.0889 | 1.0000 | 0.2159 | 0.6098 | -0.0263 | 1.0000 |

Table5d. 2d. Beta's, Standard errors and Error correlations (*N*=1,000)
(Base dataset of 10000 case, only every tenth case is used)

| N=1000 | beta's MoM | s.e. MoM | beta's cmp | s.e. cmp |
|---|---|---|---|---|
| Eq.y1: var. x1 | 0.9058 | 0.0798 | 0.8950 | 0.0820 |
| Eq.y1: var. x2 | 0.9667 | 0.0719 | 0.9531 | 0.0718 |
| Eq.y1: var. x3 | 0.8916 | 0.0726 | 0.8826 | 0.0651 |
| Eq.y1: var. x4 | 0.9127 | 0.0880 | 0.9065 | 0.0839 |
| Eq.y1: cutp. 1 | 0.0109 | 0.0581 | 0.0193 | 0.0578 |
| Eq.y2: var. x1 | 1.0706 | 0.0998 | 1.0462 | 0.0885 |
| Eq.y2: var. x2 | 1.8804 | 0.1089 | 1.8831 | 0.1162 |
| Eq.y2: var. x3 | 2.9807 | 0.1617 | 2.9809 | 0.1636 |
| Eq.y2: var. x4 | 3.9341 | 0.2122 | 3.9386 | 0.2157 |
| Eq.y2: cutp. 1 | -1.0250 | 0.0946 | -1.0196 | 0.0955 |
| Eq.y2: cutp. 2 | 1.3482 | 0.1172 | 1.3420 | 0.1055 |
| Eq.y3: var. x1 | -1.1157 | 0.1235 | -1.1258 | 0.1419 |
| Eq.y3: var. x2 | 2.1102 | 0.1724 | 2.1565 | 0.1987 |
| Eq.y3: var. x3 | -3.2868 | 0.2639 | -3.3482 | 0.2912 |
| Eq.y3: var. x4 | 4.4391 | 0.3299 | 4.5149 | 0.3847 |
| Eq.y3: cutp. 1 | -1.1548 | 0.1573 | -1.1448 | 0.1512 |
| Eq.y3: cutp. 2 | 0.5539 | 0.1232 | 0.5435 | 0.1322 |
| Eq.y4: var. x1 | -1.0006 | 0.0658 | -1.0014 | 0.0664 |
| Eq.y4: var. x2 | 0.4593 | 0.0526 | 0.4538 | 0.0507 |
| Eq.y4: var. x3 | -1.0533 | 0.0529 | -1.0494 | 0.0562 |
| Eq.y4: var. x4 | 0.9190 | 0.0742 | 0.9095 | 0.0665 |
| Eq.y4: cutp. 1 | -1.6660 | 0.0938 | -1.7127 | 0.0892 |
| Eq.y4: cutp. 2 | -0.5368 | 0.0729 | -0.5796 | 0.0724 |
| Eq.y4: cutp. 3 | 1.0163 | 0.0762 | 1.0410 | 0.0778 |

| Estimated Full correlation matrix MoM | | | | Estimated Full correlation matrix cmp | | | |
|---|---|---|---|---|---|---|---|
| 1.0000 | | | | 1.0000 | | | |
| 0.7347 | 1.0000 | | | 0.5360 | 1.0000 | | |
| -0.3041 | 0.0570 | 1.0000 | | -0.3291 | 0.2410 | 1.0000 | |
| 0.1500 | 0.6412 | -0.1904 | 1.0000 | 0.2013 | 0.5261 | -0.1288 | 1.0000 |

0.7347: 95% confidence interval cmp [0.3862 : 0.6583]

Table5c. 2e. Beta's, Standard errors and Error correlations ($N$=1,000)
(A new created dataset of 1000 cases)

| N=1000 | beta's MoM | s.e. MoM | beta's cmp | s.e. cmp |
|---|---|---|---|---|
| Eq.y1: var. x1 | 0.9490 | 0.0846 | 0.9536 | 0.0827 |
| Eq.y1: var. x2 | 0.9997 | 0.0726 | 1.0142 | 0.0747 |
| Eq.y1: var. x3 | 0.9368 | 0.0702 | 0.9446 | 0.0696 |
| Eq.y1: var. x4 | 0.9916 | 0.0871 | 0.9944 | 0.0908 |
| Eq.y1: cutp. 1 | 0.0282 | 0.0604 | 0.0380 | 0.0599 |
| Eq.y2: var. x1 | 0.8773 | 0.0956 | 0.8964 | 0.0795 |
| Eq.y2: var. x2 | 1.9966 | 0.1537 | 1.9868 | 0.1219 |
| Eq.y2: var. x3 | 2.9308 | 0.2170 | 2.9092 | 0.1597 |
| Eq.y2: var. x4 | 3.8298 | 0.2892 | 3.8065 | 0.2112 |
| Eq.y2: cutp. 1 | -0.9652 | 0.0989 | -0.9571 | 0.0952 |
| Eq.y2: cutp. 2 | 1.2944 | 0.1303 | 1.3508 | 0.1049 |
| Eq.y3: var. x1 | -1.0870 | 0.1472 | -1.0583 | 0.1484 |
| Eq.y3: var. x2 | 2.1040 | 0.2332 | 2.1304 | 0.2109 |
| Eq.y3: var. x3 | -3.2458 | 0.2952 | -3.2709 | 0.2838 |
| Eq.y3: var. x4 | 4.2512 | 0.4029 | 4.3088 | 0.3815 |
| Eq.y3: cutp. 1 | -1.0302 | 0.1601 | -1.0904 | 0.1513 |
| Eq.y3: cutp. 2 | 0.6065 | 0.1224 | 0.5906 | 0.1338 |
| Eq.y4: var. x1 | -0.9638 | 0.0650 | -0.9645 | 0.0649 |
| Eq.y4: var. x2 | 0.4585 | 0.0521 | 0.4732 | 0.0518 |
| Eq.y4: var. x3 | -0.8987 | 0.0531 | -0.9015 | 0.0500 |
| Eq.y4: var. x4 | 0.9384 | 0.0712 | 0.9204 | 0.0672 |
| Eq.y4: cutp. 1 | -1.4835 | 0.0854 | -1.4945 | 0.0842 |
| Eq.y4: cutp. 2 | -0.5339 | 0.0690 | -0.5255 | 0.0680 |
| Eq.y4: cutp. 3 | 0.8221 | 0.0821 | 0.8277 | 0.0699 |

| Estimated Full correlation matrix MoM | | | | Estimated Full correlation matrix cmp | | | |
|---|---|---|---|---|---|---|---|
| 1.0000 | | | | 1.0000 | | | |
| 0.5582 | 1.0000 | | | 0.3511 | 1.0000 | | |
| -0.5354 | 0.4227 | 1.0000 | | -0.5507 | 0.1285 | 1.0000 | |
| 0.3158 | 0.7640 | -0.3565 | 1.0000 | 0.1896 | 0.6688 | -0.1589 | 1.0000 |

0.5582: 95% confidence interval cmp [ 0.1834 : 0.4989]
0.4227: 95% confidence interval cmp [-0.1869 : 0.4199]

Table5c.2f. Beta's, Standard errors and Error correlations (*N*=1,000)
(Again a new created dataset of 1000 cases)

|  | beta's MoM | s.e. MoM | beta's cmp | s.e. cmp |
|---|---|---|---|---|
| Eq.y1: var. x1 | 0.9592 | 0.0809 | 0.9417 | 0.0800 |
| Eq.y1: var. x2 | 0.9477 | 0.0738 | 0.9352 | 0.0708 |
| Eq.y1: var. x3 | 0.9064 | 0.0648 | 0.8982 | 0.0652 |
| Eq.y1: var. x4 | 0.9397 | 0.0831 | 0.9374 | 0.0833 |
| Eq.y1: cutp. 1 | 0.0766 | 0.0588 | 0.0701 | 0.0577 |
| Eq.y2: var. x1 | 0.9359 | 0.0837 | 0.9287 | 0.0860 |
| Eq.y2: var. x2 | 2.0244 | 0.1159 | 2.0294 | 0.1236 |
| Eq.y2: var. x3 | 2.8800 | 0.1441 | 2.8982 | 0.1618 |
| Eq.y2: var. x4 | 3.8478 | 0.1961 | 3.8657 | 0.2161 |
| Eq.y2: cutp. 1 | -0.9785 | 0.0996 | -0.9942 | 0.0962 |
| Eq.y2: cutp. 2 | 1.4898 | 0.1128 | 1.4802 | 0.1161 |
| Eq.y3: var. x1 | -0.8380 | 0.1291 | -0.8675 | 0.1328 |
| Eq.y3: var. x2 | 1.8184 | 0.1703 | 1.8429 | 0.1721 |
| Eq.y3: var. x3 | -2.8861 | 0.2167 | -2.8964 | 0.2404 |
| Eq.y3: var. x4 | 3.8474 | 0.2721 | 3.8491 | 0.3136 |
| Eq.y3: cutp. 1 | -1.0368 | 0.1240 | -1.0327 | 0.1390 |
| Eq.y3: cutp. 2 | 0.3357 | 0.1239 | 0.3267 | 0.1197 |
| Eq.y4: var. x1 | -1.0838 | 0.0679 | -1.0809 | 0.0705 |
| Eq.y4: var. x2 | 0.5400 | 0.0531 | 0.5334 | 0.0530 |
| Eq.y4: var. x3 | -1.0304 | 0.0550 | -1.0243 | 0.0558 |
| Eq.y4: var. x4 | 1.0777 | 0.0685 | 1.0711 | 0.0698 |
| Eq.y4: cutp. 1 | -1.5268 | 0.0823 | -1.5160 | 0.0856 |
| Eq.y4: cutp. 2 | -0.4549 | 0.0726 | -0.4652 | 0.0695 |
| Eq.y4: cutp. 3 | 1.0477 | 0.0764 | 1.0275 | 0.0767 |

| Estimated Full correlation matrix MoM | | | | Estimated Full correlation matrix cmp | | | |
|---|---|---|---|---|---|---|---|
| 1.0000 | | | | 1.0000 | | | |
| 0.6303 | 1.0000 | | | 0.6316 | 1.0000 | | |
| -0.1647 | 0.5655 | 1.0000 | | -0.1230 | 0.2274 | 1.0000 | |
| 0.1451 | 0.6184 | 0.0078 | 1.0000 | 0.1891 | 0.5982 | 0.0324 | 1.0000 |

(0.5655: 95% confidence interval cmp [-0.1068 : 0.5155)

We conclude that the method is stable in the number *N of observations* and it does not differ significantly from the `cmp`-estimates.

## 6. Concluding remarks.

In this paper we suggest a new approach to the statistical analysis of ordinal data, where the errors are supposed to be correlated. The basic idea is that the ordinally observed dependent variables reflect latent continuously-valued random variables $Y$ and that the observations are coarsened corresponding to an interval grid $\mathbb{C} = \left\{ \left( \nu_{j-1}, \nu_j \right] \right\}_{j=1}^{J}$ of $Y$ on the real axis, where the unknown cut-points $\nu$ have to be estimated as well. Each observed category corresponds to one interval on the real axis. For cases where $Y$ is more-dimensional, say $K$, and errors are correlated there is mostly a formidable impediment. The usual ML-estimation procedure requires to evaluate likelihoods, which are multivariate normal integrals over $K$-dimensional blocks. If this has to be performed this is very cumbersome and time-consuming. In this paper we show that estimation of the latent generating model behind the coarsened observations of dependent variables can be done in a much simpler way than usual without the need for multi-dimensional integration or large-scale simulations.

In our approach we depart from the requirement that the difference between the observation $Y$ and its predictor $\beta' X$, that is $(Y - \beta' X)$, cannot be further explained by $X$. This is translated into the zero-covariance conditions (2.5) and gives those conditions a significance on their own. When the errors are normally distributed this coincides with the ML-conditions.

The identifying moment conditions are found by substituting the residuals in the regular zero covariance-conditions for exact data by the corresponding generalized residuals corresponding to the ordinal data.

The approach closely resembles the traditional GLS- and SUR-approaches used to estimate linear models on exactly observed dependent variables.

For this method an assumption about the marginal distributions of the error vector is required. We choose for normality, which enables us to use (3.6). Although we restricted ourselves to assuming normal errors, it is not difficult to generalize this method for other error distributions as well, where the logistic and the lognormal are the foremost candidates (see in those cases, e.g., Maddala (1983, p.369) for the formulae of the generalized residuals). The estimation method remains unchanged.

This approach seems to smoothly close the gap between the analysis on exactly observable data and qualitative ordinal data. We saw in the above examples that the effect on variances (confidence bands and intervals) caused by SUOP compared to OP is in some cases small and in other cases large. The regression coefficients are mostly similar, which is not surprising as both estimators are consistent estimators. This is also the case for the comparison between traditional OLS and SUR estimates in traditional econometrics. The advantage lies in the possibility to account for error correlations, caused by using the additional information supplied by the error correlations. Standard error deviations are assessed without assuming a specific structure of the covariance matrix before estimation. In the panel data example in Section 6a it appears that the standard deviations of the estimates are doubled or more taking error correlation into account. Hence, in this case the reliability reduction when taking error correlations into account is huge.

In this paper we focus on the qualitative versions FMOP and SUOP of FGLS and SUR. However, this method seems generally appropriate for two broad types of model estimation situations characterized by Roodman (2011) as:

"1)      those in which a truly recursive data-generating process is posited and fully modeled, and

2) those in which there is simultaneity but instruments allow the construction of a recursive set of equations, as in two-stage least squares (2SLS)."

Our method may be compared with the methods, (based on the GHK algorithm), developed by Capellari and Jenkins (2003), based on simulated moments (Hajivassiliou and McFadden (1998), and Roodman (2007, 2020)). Those methods aim at getting numerical estimates of the log-likelihood by simulation and finding, by variation of the unknown parameters to be estimated, which parameter values maximize the simulated log-likelihood of the sample. This requires the repeated evaluation of multiple normal integrals and makes the procedure time-consuming. In our approach we do not need to evaluate multiple integrals nor large-scale simulations, and therefore the method is not restricted with respect to the size $K$ of the equation system. Moreover, we can handle an arbitrary number $J(>2)$ of outcome categories. We do not have to restrict ourselves to dichotomous (biprobit) data only. The method can be used for any number of equations $K$ and any number of interval categories $J_k$. For instance, in our SUOP-example (Section 6.b) we estimated eight equations, 75 effects and 32 cut points simultaneously. It is obvious that direct observation is a limiting case of coarsening and consequently the methos may also be used when the data set consists of a mixture of directly observed and ordinal data. Classical least-squares based estimation methods on exactly observed data may be seen as a specific limiting case.

Our estimation method appears to require only a few minutes of computing time, which compares favorably with the traditional methods each of which requires much more time. The method may be interpreted as a generalization of classical least squares models that deal with exact observations to include the estimation of models on the basis of more-dimensional ordered probit-type observations.

In this paper we restricted ourselves to the most straightforward OP observation mode. In a forthcoming study we will generalize this approach to tackle the case where the sample consists of a mixture of categorical, censored, and exactly observed data.

**Appendix.**

**A.** **Estimation of the latent error covariance matrix Σ.**

For the estimation of the latent covariance matrix starting from the estimated $\hat{b}$-estimates and the in-between covariance matrix $\hat{\bar{\Sigma}}$ we propose the following method. We make use of the fact that in order to estimate the true error covariances $\rho_{k,k'}$ by $\hat{\rho}_{k,k'}$ for two OP-equations we notice that for each pair $k,k'$ we only have to look for the bivariate marginal distribution of $\varepsilon_k, \varepsilon_{k'}$. Consider for instance a cluster of three observations (1, 2, 3). The sample likelihood is $P(\varepsilon_1 \in S_1,\ \varepsilon_2 \in S_2,\ \varepsilon_3 \in S_3\ ;0,\Sigma_{1,2,3})$, where $S_1 = (\nu_{j_i-1} - x_{i,1}\beta < \varepsilon_{i,1} \le \nu_{j_i} - x_{i,1}\beta]$ and $S_2$ and $S_3$ similarly defined. However, the bivariate marginal likelihood provides the same information on $\sigma_{1,2}$ as the trivariate full likelihood. In the $K$-dimensional covariance matrix there are $K(K-1)/2$ different non-diagonal elements For instance, if $K = 8$ as in the model in Section 6, this boils down to 28 simple two-dimensional estimations.

Applying the Gram-Schmidt decomposition, there holds for the pair $\varepsilon_k, \varepsilon_{k'}$

$$\varepsilon_{i,k'} = \rho_{k,k'}.\varepsilon_{i,k} + \eta_{i,k}\sqrt{1-\rho_{k,k'}^2} \tag{A.1}$$

where $\varepsilon_{i,k}, \eta_{i,k}$ are independent $N(0,1)$ drawings.[9] The coefficient $\sqrt{1-\rho_{k,k'}^2}$ guarantees that $\sigma^2(\varepsilon_{i,k'}) = 1$. We have $E(\varepsilon_{i,k}.\varepsilon_{i,k'}) = \rho_{k,k'}$. When we consider the estimated in-between error covariance $\bar{\rho}_{k,k'} = E(\bar{\varepsilon}_{i,k}, \bar{\varepsilon}_{i,k'})$, then its value depends only on the two one-dimensional grids $\mathbb{C}_{ik}$ and $\mathbb{C}_{ik'}$ and $\rho_{k,k'}$. We may now simulate $\varepsilon_{i,k}, \varepsilon_{i,k'}$ for various values of $\rho_{k,k'}$ by means of (A.1), and calculate the corresponding in-between covariances $\bar{\rho}_{k,k'} = \frac{1}{N}\sum_{i=1}^{N}\bar{e}_{i,k}.\bar{e}_{i,k'}$, corresponding to the grids $\mathbb{C}_{ik}$ and $\mathbb{C}_{ik'}$. We notice that the dependent variable vector $Y_{i,k}$ is observed according to a uniform $i$-independent grid $\mathbb{C}_k = \left\{(\nu_{j-1}^{(k)}, \nu_j^{(k)}]\right\}_{j=1}^{J_k} = \left\{\mathbb{S}_j^{(k)}\right\}_{j=1}^{J_k}$, while the errors $\varepsilon_{i,k}$ are observed according to $i$-dependent individual grids

$$\mathbb{C}_{ik} = \left\{(\nu_{j-1}^{(k)} - x_{i,k}\beta,\ \nu_j^{(k)} - x_{i,k}\beta]\right\}_{j=1}^{J_k} = \left\{\mathbb{S}_j^{(k)} - x_{i,k}\beta\right\}_{j=1}^{J_k}.$$

We consider the set of all two-dimensional grids for all observation units. $\left\{\mathbb{C}_{ik}\ ,\ \mathbb{C}_{ik'}\right\}_{i=1}^{N} = \mathbb{C}_{k,k'}$

We write for the sample in-between covariance

$$\bar{\rho}_{k,k'} = \frac{1}{N}\sum_{i=1}^{N}\bar{e}_{i,k}.(\rho_{k,k'}).\bar{e}_{i,k'}.(\rho_{k,k'}) \overset{def}{=} f(\rho_{k,k'}).$$

We estimate the value of the latent $\rho_{k,k'}$ by comparing the observed sample in-between covariance $\hat{\bar{\rho}}_{k,k'}$ with its simulated counterpart $\bar{\rho}_{k,k'}$ for various values of the latent $\rho_{k,k'}$. It appears that there is one value $\hat{\rho}_{k,k'}$ solving $f(\rho_{k,k'}) = \hat{\bar{\rho}}_{k,k'}$. In order to gain insight into the relation between the latent $\rho_{k,k'}$ and the corresponding in-between covariance $\bar{\rho}_{k,k'}$ we did some simulation experiments for three different two-dimensional grids. In Table A1 we present three different two-dimensional grids with the different $\bar{\rho}_{k,k'}$ values for different values of $\rho_{k,k'}$. We describe one example in detail. Consider the two-dimensional grid with $\mathbb{C}_1 = \{(-\infty,0],(0,\infty)\}$, $\mathbb{C}_2 = \{(-\infty,0],(0,\infty)\}$ in the middle part of Table A1. We simulate a sample from a two-dimensional normal distribution with a correlation with $\rho$ =0.1. We find an in-between covariance $\bar{\rho}$ =0.043. For $\rho$ =0.2 we find a corresponding value of $\bar{\rho}$ =0.087, and so on. In Table A1 we present the relation between $\rho$ and $\bar{\rho}$ for three different two-dimensional grids. We found that the function $f(\rho_{k,k'})$ = $\bar{\rho}_{k,k'}$ is monotonically increasing in $\rho_{k,k'}$ for all grids we tested. See also Aitkin(1964). We conclude that for a given grid $(\mathbb{C}_1, \mathbb{C}_2)$ the function $f(\rho) = \bar{\rho}$ is monotonically increasing in $\rho$.

---

[9] We use standard normal draws for $\varepsilon_{i,k},\ \eta_{i,k'}$, because in (A.1) the simulated errors have to be $N(0,1)$.

Table A1. Relation between covariance $\rho$ and in-between covariance $\bar{\rho}$.

| grid | $\rho$ | $\bar{\rho}$ | grid | $\rho$ | $\bar{\rho}$ | grid | $\rho$ | $\bar{\rho}$ |
|---|---|---|---|---|---|---|---|---|
| $v_{11}$=-0.5 | 0.1 | 0.075 | $v_1$=0.0 | 0.1 | 0.043 | $v_{11}$=-1.0 | 0.1 | 0.032 |
| $v_{12}$= 0.0 | 0.2 | 0.141 | | 0.2 | 0.087 | $v_{12}$= 2.0 | 0.2 | 0.051 |
| $v_{13}$= 0.75 | 0.3 | 0.198 | | 0.3 | 0.128 | | 0.3 | 0.092 |
| | 0.4 | 0.290 | $v_2$=0.0 | 0.4 | 0.167 | $v_{21}$=-2.0 | 0.4 | 0.119 |
| $v_{21}$=-0.75 | 0.5 | 0.342 | | 0.5 | 0.224 | $v_{22}$= 1.0 | 0.5 | 0.140 |
| $v_{22}$=-0.5 | 0.6 | 0.451 | | 0.6 | 0.264 | | 0.6 | 0.167 |
| $v_{23}$= 0.5 | 0.7 | 0.505 | | 0.7 | 0.321 | | 0.7 | 0.189 |
| N=10,000 | 0.8 | 0.601 | N=10,000 | 0.8 | 0.367 | N=10,000 | 0.8 | 0.217 |
| | 0.9 | 0.681 | | 0.9 | 0.457 | | 0.9 | 0.219 |

The solution $\hat{\rho}_{k,k'}$ of $f(\rho_{k,k'}) = \hat{\bar{\rho}}_{k,k'}$ is, using Slutsky's Law, a consistent estimator $\hat{\rho}_{k,k'}$ of the population parameter $\rho_{k,k'}$. Doing this for each non-diagonal element of $\Sigma$, we estimate the underlying non-diagonal elements of the full correlation matrix $\Sigma$. The diagonal elements are equal to one by assumption. This yields a consistent estimator $\hat{\Sigma}_\varepsilon$ of the error covariance matrix $\Sigma_\varepsilon$. Confidence intervals can be found by the delta-method. Fig. A1 shows the graph of the function $f(\rho_{k,k'})$ for the left grid in Table A1.

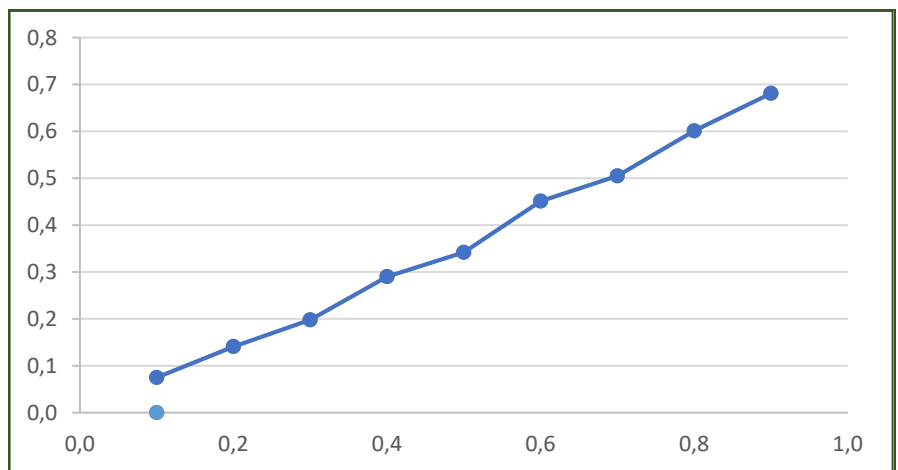

Fig. A1. $\overline{\rho}$ as a function of $\rho$

In Table 4 we presented the full correlation matrices calculated by SUOP and by `cmp`. In Table A2. below we present the in-between covariance matrix and the full covariance matrix for the five-year panel considered in Section 6a. We notice the fact that the grid-wise observation causes a considerable loss of information. About 60% of the information is lost. The estimates are not based on structural assumptions like 'random effects' or errors where the correlations are specific functions of the time difference. Actually, this result could be used to estimate and test specific functional specifications of the covariance matrix. From table A2. it is already obvious that 'random effects' is to be rejected for this panel structure, because in that case the non-diagonal elements would have to be roughly equal to each other.

Table A2. In-between and full covariance matrices for the five-year panel

|  | In-between covariance matrix | | | | | Full covariance matrix | | | | |
|---|---|---|---|---|---|---|---|---|---|---|
|  | 2009 | 2010 | 2011 | 2012 | 2013 | 2009 | 2010 | 2011 | 2012 | 2013 |
| 2009 | .6434 |  |  |  |  | 1.0000 |  |  |  |  |
| 2010 | .4852 | .6292 |  |  |  | .8899 | 1.0000 |  |  |  |
| 2011 | .4123 | .4841 | .6289 |  |  | .7984 | .8945 | 1.0000 |  |  |
| 2012 | .3771 | .4210 | .4793 | .6277 |  | .7490 | .8133 | .8910 | 1.0000 |  |
| 2013 | .3413 | .3725 | .4156 | .4944 | .6321 | .7064 | .7482 | .7970 | .9270 | 1.0000 |

## B.    Table 3 complete.

Table 3. Comparison of the parameter estimates and their s.e.'s for Ordered Probit, Method of Moments, and Maximum Likelihood.

| McK-Z $R^2$ | 0.1489 | | 0.1481 | | 0.1154 | |
|---|---|---|---|---|---|---|
| *Health Satisfaction* | $\beta_{OP}$ | $\sigma_{OP}$ | $\beta_{SUOP}$ | $\sigma_{SUOP}$ | $\beta_{ML}$ | $\sigma_{ML}$ |
| Health: AGE | -.0695 | .0046 | -.0693 | .0048 | -.0651 | .0045 |
| Health: AGE$^2$ | .0006 | .00005 | .0006 | .00005 | .0006 | .00005 |
| Health: D_FEMALE | -.0182 | .0175 | -.0181 | .0174 | -.0134 | .0166 |
| Health: D_SINGLE | -.0514 | .0291 | -.0510 | .0293 | -.0570 | .0278 |
| Health: D_SEPARATED | -.1241 | .0240 | -.1238 | .0243 | -.1108 | .0230 |
| Health: Ln_LABOURINC | .0289 | .0025 | .0288 | .0026 | .0227 | .0022 |
| Health: CHILDREN | .0357 | .0123 | .0356 | .0126 | .0308 | .0114 |
| Health: D_EAST | -.1678 | .0201 | -.1670 | .0196 | -.1402 | .0193 |
| Health: DISABLE | -.7991 | .0272 | -.7971 | .0268 | -.5825 | .0232 |
| Health: Cut point 1 | -1.8277 | .0184 | -1.8240 | .0191 | -1.8163 | .0181 |
| Health: Cut point 2 | -1.1161 | .0131 | -1.1094 | .0134 | -1.1122 | .0130 |
| Health: Cut point 3 | -.3432 | .0109 | -.3415 | .0109 | -.3399 | .0107 |
| Health: Cut point 4 | .9468 | .0123 | .9442 | .0124 | .9409 | .0122 |
| McK-Z $R^2$ | 0.0278 | | 0.0278 | | 0.0196 | |
| *Sleep Satisfaction* | $\beta_{OP}$ | $\sigma_{OP}$ | $\beta_{SUOP}$ | $\sigma_{SUOP}$ | $\beta_{ML}$ | $\sigma_{ML}$ |
| Sleep: AGE | -.0380 | .0041 | -.0381 | .0042 | -.0348 | .0041 |
| Sleep: AGE$^2$ | .0004 | .00004 | .0004 | .00004 | .0003 | .00004 |
| Sleep: D. FEMALE | -.1213 | .0172 | -.1213 | .0172 | -.1204 | .0164 |
| Sleep: D. SINGLE | .0508 | .0302 | .0505 | .0307 | .0594 | .0284 |
| Sleep: D. SEPARATED | -.0762 | .0253 | -.0762 | .0260 | -.0658 | .0238 |
| Sleep: Ln(HH.LABOURINC) | .0067 | .0021 | .0066 | .0020 | .0062 | .0018 |
| Sleep: # of CHILDREN | .0219 | .0121 | .0220 | .0122 | .0164 | .0113 |
| Sleep: YEARS of EDUCATION | .0316 | .0032 | .0315 | .0031 | .0117 | .0028 |
| Sleep: D. EAST GERMANY | -.1063 | .0199 | -.1065 | .0196 | -.0763 | .0192 |
| Sleep: Cut point 1 | -1.6918 | .0175 | -1.6912 | .0174 | -1.7036 | .0174 |
| Sleep: Cut point 2 | -.9791 | .0123 | -.9800 | .0123 | -.9831 | .0123 |
| Sleep: Cut point 3 | -.9251 | .0106 | -.2970 | .0106 | -.2940 | .0104 |
| Sleep: Cut point 4 | .7286 | .0114 | .7269 | .0115 | .7150 | .0113 |

| McK-Z $R^2$ | 0.1345 | | 0.1342 | | 0.1133 | |
|---|---|---|---|---|---|---|
| *Household Income Satisfaction* | $\beta_{OP}$ | $\sigma_{OP}$ | $\beta_{SUOP}$ | $\sigma_{SUOP}$ | $\beta_{ML}$ | $\sigma_{ML}$ |
| HH inc.: AGE | -.0648 | .0045 | -.0649 | .0048 | -.0593 | .0045 |
| HH inc.: AGE² | .0008 | .00005 | .0008 | .00005 | .0007 | .00005 |
| HH inc.: D. FEMALE | .0837 | .0175 | .0836 | .0175 | .0764 | .0167 |
| HH inc.: D. SINGLE | -.1398 | .0304 | -.1404 | .0308 | -.1326 | .0279 |
| HH inc.: D. SEPARATED | -.2582 | .0255 | -.2583 | .0264 | -.2404 | .0237 |
| HH inc.: Ln(LABOURINC) | .0396 | .0026 | .0395 | .0027 | .0349 | .0025 |
| HH inc.: Ln(HH.LABOURINC) | .0261 | .0021 | .0261 | .0021 | .0262 | .0014 |
| HH inc.: # of CHILDREN | -.0088 | .0122 | -.0085 | .0122 | -.0212 | .0095 |
| HH inc.: YEARS of EDUCATION | .0808 | .0033 | .0806 | .0034 | .0693 | .0032 |
| HH inc.: D. EAST GERMANY | -.3862 | .0201 | -.3867 | .0200 | -.3472 | .0185 |
| HH-inc.: Cut point 1 | -1.7878 | .0179 | -1.7106 | .0174 | -1.7012 | .0175 |
| HH-inc.:Cut point 2 | -1.0544 | .0129 | -1.0564 | .0128 | -1.0323 | .0126 |
| HH-inc.:Cut point 3 | -.2814 | .0108 | -.2858 | .0107 | -.2743 | .0103 |
| HH-inc.:Cut point 4 | .9480 | .0123 | .9442 | .0126 | .9020 | .0121 |
| McK-Z $R^2$ | 0.1503 | | 0.1495 | | 0.1306 | |
| *Individual Income Satisfaction* | $\beta_{OP}$ | $\sigma_{OP}$ | $\beta_{SUOP}$ | $\sigma_{SUOP}$ | $\beta_{ML}$ | $\sigma_{ML}$ |
| Ind. inc.: AGE | -.0562 | .0045 | -.0562 | .0048 | -.0565 | .0045 |
| Ind. inc.: AGE² | .0008 | .00005 | .0008 | .00005 | .0007 | .00005 |
| Ind. inc.: D. FEMALE | -.1239 | .0174 | -.1239 | .0173 | -.1180 | .0169 |
| Ind. inc.: D. SINGLE | -.0595 | .0290 | -.0602 | .0292 | -.0613 | .0277 |
| Ind. inc.: D. SEPARATED | -.1016 | .0240 | -.1016 | .0245 | -.0885 | .0233 |
| Ind. inc.: Ln(LABOURINC) | .0727 | .0026 | .0725 | .0028 | .0728 | .0025 |
| Ind. inc.: # of CHILDREN | .0311 | .0121 | .0314 | .0119 | .0277 | .0102 |
| Ind. inc.: YEARS of EDUCATION | .0752 | .0033 | .0749 | .0034 | .0670 | .0032 |
| Ind. inc.: D. EAST GERMANY | -.2912 | .0200 | -.2916 | .0198 | -.2561 | .0189 |
| Ind. inc.: DISABLLILITY RATE | -.1510 | .0270 | -.1513 | .0272 | .0430 | .0177 |
| Ind. Inc.: Cut point 1 | -1.3425 | .0147 | -1.3428 | .0144 | -1.2938 | .0141 |
| Ind. inc.: Cut point 2 | -.7372 | .0117 | -.7401 | .0116 | -.7361 | .0114 |
| Ind. inc.: Cut point 3 | -.0300 | .0107 | -.0358 | .0106 | -.0442 | 0103 |
| Ind. inc.: Cut point 4 | 1.1096 | .0129 | 1.1043 | .0133 | 1.0914 | .0130 |
| McK-Z $R^2$ | 0.0566 | | 0.0564 | | 0.0475 | |
| *Dwelling Satisfaction* | $\beta_{OP}$ | $\sigma_{OP}$ | $\beta_{SUOP}$ | $\sigma_{SUOP}$ | $\beta_{ML}$ | $\sigma_{ML}$ |
| Dwelling: AGE | -.0432 | .0047 | -.0424 | .0049 | -.0397 | .0046 |
| Dwelling: AGE² | .0006 | .00005 | .0006 | .00005 | .0005 | .00005 |
| Dwelling: D. FEMALE | .0333 | .0183 | .0332 | .0182 | .0409 | .0168 |
| Dwelling: D. SINGLE | -.2409 | .0313 | -.2414 | .0322 | -.2312 | .0282 |
| Dwelling: D. SEPARATED | -.2384 | .0265 | -.2384 | .0266 | -.2126 | .0242 |
| Dwelling: Ln(LABOURINC) | .0179 | .0027 | .0178 | .0027 | .0151 | .0026 |
| Dwelling: Ln(HH.LABOURINC) | .0134 | .0022 | .0133 | .0022 | .0142 | .0020 |
| Dwelling: # of CHILDREN | -.0079 | .0127 | -.0077 | .0127 | -.0159 | .0109 |
| Dwelling: YEARS of EDUC. | .0313 | .0034 | .0312 | .0034 | .0213 | .0033 |
| Dwelling: DISABLLILITY RATE | -.1317 | .0283 | -.1320 | .0287 | .0258 | .0245 |
| Dwelling: Cut point 1 | -2.1862 | .0261 | -2.1806 | .0256 | -2.1637 | .0252 |
| Dwelling: Cut point 2 | -1.6332 | .0172 | -1.6321 | .0170 | -1.6211 | .0169 |
| Dwelling: Cut point 3 | -.9408 | .0123 | -.9453 | .0122 | -.9364 | 0121 |
| Dwelling: Cut point 4 | .2504 | .0106 | .2477 | .0107 | .2360 | .0105 |

| McK-Z $R^2$ | 0.0902 | | 0.0901 | | 0.0940 | |
|---|---|---|---|---|---|---|
| *Leisure Time Satisfaction* | $\beta_{OP}$ | $\sigma_{OP}$ | $\beta_{SUOP}$ | $\sigma_{SUOP}$ | $\beta_{ML}$ | $\sigma_{ML}$ |
| Leisure: AGE | -.0433 | .0043 | -.0432 | .0044 | -.0406 | .0042 |
| Leisure: AGE$^2$ | .0006 | .00005 | .0006 | .00005 | .0005 | .00005 |
| Leisure: D. FEMALE | -.0353 | .0175 | -.0352 | .0175 | -.0257 | .0160 |
| Leisure: Ln(LABORINC) | -.0273 | .0026 | -.0272 | .0028 | -.0299 | .0025 |
| Leisure: Ln(HH.LABOURINC) | .0047 | .0019 | .0047 | .0020 | .0051 | .0018 |
| Leisure: # of CHILDREN | -.0847 | .0114 | -.0848 | .0116 | -.0896 | .0102 |
| Leisure: YEARS of EDUCATION | .0055 | .0033 | -.0056 | .0032 | -.0043 | .0031 |
| Leisure: D. EAST GERMANY | -.1293 | .0202 | -.1288 | .0200 | -.0921 | .0185 |
| Leisure: DISABLE RATE | -.1039 | .0277 | -.1036 | .0287 | .0498 | .0249 |
| Leisure: Cut point 1 | -1.8463 | .0194 | -1.8454 | .0200 | -1.8090 | .0188 |
| Leisure: Cut point 2 | -1.2129 | .0136 | -1.2092 | .0138 | -1.2051 | .0133 |
| Leisure: Cut point 3 | -.4716 | .0109 | -.4687 | .0110 | -.4846 | .0108 |
| Leisure: Cut point 4 | .6486 | .0113 | .6499 | .0113 | .6359 | .0114 |
| McK-Z $R^2$ | 0.0834 | | 0.0834 | | 0.0763 | |
| *Family Life Satisfaction* | $\beta_{OP}$ | $\sigma_{OP}$ | $\beta_{SUOP}$ | $\sigma_{SUOP}$ | $\beta_{ML}$ | $\sigma_{ML}$ |
| Family life: AGE | -.0572 | .0048 | -.0573 | .0048 | -.0534 | .0046 |
| Family life: AGE$^2$ | .0006 | .00005 | .0006 | .00005 | .0005 | .00005 |
| Family life: D. SINGLE | -.4871 | .0297 | -.4880 | .0295 | -.4665 | .0270 |
| Family life: D. SEPARATED | -.5665 | .0260 | -.5666 | .0267 | -.5420 | .0236 |
| Family life: Ln(LABORINC) | .0013 | .0027 | .0012 | .0028 | -.0012 | .0025 |
| Fam. life: Ln(HH.LABOURINC) | .0167 | .0022 | .0166 | .0022 | .0180 | .0020 |
| Family life: YEARS of EDUC. | .0040 | .0034 | .0038 | .0034 | -.0084 | .0032 |
| Family life: D. EAST GERMANY | -.1370 | .0209 | -.1376 | .0208 | -.1149 | .0192 |
| Family life: DISABLE RATE | -.1290 | .0281 | -.1293 | .0289 | .0321 | .0239 |
| Family life: Cut point 1 | -2.1949 | .0250 | -2.1905 | .0244 | -2.1609 | .0240 |
| Family life: Cut point 2 | -1.6631 | .0172 | -1.6638 | .0169 | -1.6476 | .0168 |
| Family life: Cut point 3 | -.9763 | .0125 | -.9813 | .0124 | -.9741 | .0123 |
| Family life: Cut point 4 | .1778 | .0106 | .1749 | .0107 | .1609 | .0105 |
| McK-Z $R^2$ | 0.1193 | | 0.1187 | | 0.0965 | |
| *Standard of Living Satisfaction* | $\beta_{OP}$ | $\sigma_{OP}$ | $\beta_{SUOP}$ | $\sigma_{SUOP}$ | $\beta_{ML}$ | $\sigma_{ML}$ |
| Stand.living: AGE | -.0760 | .0047 | -.0758 | .0048 | -.0699 | .0045 |
| Stand.living: AGE$^2$ | .0008 | .00005 | .0008 | .00005 | .0008 | .00005 |
| Stand.living: D. FEMALE | .1111 | .0180 | .1108 | .0180 | .1113 | .0155 |
| Stand.living: D. SINGLE | -.2982 | .0293 | -.2984 | .0292 | -.2762 | .0266 |
| Stand.living: D. SEPARATED | -.3720 | .0259 | -.3712 | .0265 | -.3415 | .0234 |
| Stand.living: Ln(LABOURINC) | .0312 | .0026 | .0310 | .0027 | .0269 | .0025 |
| Std.living: Ln(HH.LABOURINC) | .0225 | .0021 | .0225 | .0022 | .0228 | .0017 |
| Stand.living: YEARS of EDUC. | .0714 | .0034 | .0711 | .0034 | .0576 | .0032 |
| Stand.living: D. EAST GERM. | -.3197 | .0205 | -.3199 | .0201 | -.2755 | .0180 |
| Stand.living: Cut point 1 | -2.2120 | .0254 | -2.1991 | .0251 | -2.2096 | .0242 |
| Stand.living: Cut point 2 | -1.6048 | .0167 | -1.6018 | .0164 | -1.5980 | .0163 |
| Stand.living: Cut point 3 | -.8760 | .0119 | -.8219 | .0118 | -.8087 | .0117 |
| Stand.living: Cut point 4 | .5312 | .0111 | .5264 | .0113 | .5095 | .0109 |